%% file: 000main.tex
\documentclass{article} %
\usepackage{arxiv}
\usepackage[utf8]{inputenc} %
\usepackage[T1]{fontenc}    %
\usepackage{url}            %
\usepackage{booktabs}       %
\usepackage{amsfonts}       %
\usepackage{nicefrac}       %
\usepackage{microtype}      %
\usepackage{lipsum}

\usepackage{url}

\usepackage{breakurl}
\usepackage[breaklinks]{hyperref}
\usepackage{savesym}
\usepackage{caption}
\usepackage{subcaption}
\usepackage{tikz}
\usepackage{pgfplots}
\usetikzlibrary{shapes,shapes.geometric,arrows,shapes.symbols,positioning,calc,pgfplots.groupplots}
\usepgfplotslibrary{groupplots}
\pgfplotsset{compat=1.16}
\usepackage{picture}
\usepackage{svg}
\usepackage{amsmath}
\usepackage{amsfonts}
\usepackage{amssymb}
\usepackage{mathtools}
\usepackage{paralist}
\usepackage{amsthm}

\savesymbol{labelindent}
\usepackage[inline]{enumitem}

\graphicspath{ {./assets/} } 
\usepackage{graphicx}
\usepackage{multirow}
\usepackage{makecell}
\usepackage{breqn}
\usepackage[ruled,linesnumbered,lined,boxed,commentsnumbered]{algorithm2e}
\usepackage{array}
\savesymbol{appendices}
\usepackage[toc,page]{appendix}
\usepackage{adjustbox}
\usepackage{xparse}
\usepackage{cleveref}

\crefformat{subfigure}{#2\onlyletter{#1}#3}
\crefrangeformat{subfigure}{(#3\onlyletter{#1}#4--#5\onlyletter{#2}#6)}
\crefmultiformat{subfigure}
  {(#2\onlyletter{#1}#3}
  {,~#2\onlyletter{#1}#3)}
  {,~#2\onlyletter{#1}#3)}
  {,~#2\onlyletter{#1}#3)}

\ExplSyntaxOn
\NewDocumentCommand{\onlyletter}{m}
 {
  \tl_set:Nx \l_tmpa_tl { #1 }
  \tl_item:Nn \l_tmpa_tl { -1 }
 }
\ExplSyntaxOff

\author{%
Aref Asvadishirehjini \\
Department of Computer Science \\
University of Texas at Dallas \\
\texttt{aref@utdallas.edu} \\
\And
Murat Kantarcioglu \\
Department of Computer Science \\
University of Texas at Dallas \\
\texttt{muratk@utdallas.edu} \\
\And
Bradley Malin \\
Department of Biomedical Informatics \\ Vanderbilt University Medical Center \\
\texttt{b.malin@vanderbilt.edu} \\
}

\usepackage{xspace}

\newcommand{\systemGOAT}{GOAT}
\newcommand{\systemGOATTitle}{GOAT}
\newcommand{\systemGOATTitleDef}{\underline{G}PU \underline{O}utsourcing of Deep Learning Training With \underline{A}synchronous \mbox{Probabilistic} Integrity Verification Inside \underline{T}rusted Execution \mbox{Environment}}

\newcommand{\systemtitle}{\systemGOATTitle{}\xspace}
\newcommand{\systemname}{\systemGOAT{}\xspace}
\newcommand{\systitledef}{{\systemGOATTitleDef{}}\xspace}

\theoremstyle{remark}
\newtheorem*{remark}{Remark}

\theoremstyle{definition}
\newtheorem{definition}{Definition}

\theoremstyle{plain}
\newtheorem{theorem}{Theorem}

\definecolor{light-gray}{gray}{0.95}

\title{\Large \bf \systemtitle{}:
	\systitledef{}}

\date{} %

\DeclarePairedDelimiter{\ceil}{\lceil}{\rceil}

\begin{document}

\maketitle

\begin{abstract}
	\input{001abst.tex}
\end{abstract}

\section{Introduction}
\label{sec:intro}
\input{002intro.tex}

\section{Background}
\label{sec:back}
\input{003back.tex}

\section{Threat Model}
\label{sec:threat}
\input{004threat.tex}

\section{System Design}
\label{sec:system}
\input{005sys.tex}

\section{Integrity Analysis}
\label{sec:integ}
\input{006sec.tex}

\section{Experimental Evaluation}
\label{sec:exp}
\input{007eval.tex}

\section{Related Works}
\label{sec:rel}
\input{008relate.tex}

\section{Conclusion}
\label{sec:conc}
\input{099concl.tex}

\subsubsection*{Acknowledgments}
This work was supported, in part by grant 2R01HG006844 from the National Human Genome Research Institute, NSF Awards CNS-1837627, OAC-1828467, IIS-1939728 and ARO award W911NF-17-1-0356. Finally, the authors would like to thank \mbox{Dr.~Yan~Zhou for} constructive criticism of the manuscript.

\bibliography{references.bib}
\bibliographystyle{unsrt}

\appendix
\input{0091adx.tex}

\end{document}

%% file: 001abst.tex
Machine learning models based on Deep Neural Networks (DNNs) are increasingly deployed in a wide range of applications, ranging from self-driving cars to COVID-19 treatment discovery. To support the computational power necessary to learn a DNN, cloud environments with dedicated hardware support have emerged as critical infrastructure. 
However, there are many 
integrity 
challenges associated with outsourcing computation.
Various approaches have been developed to address these challenges, building on 
trusted execution environments (TEE). Yet, no existing approach scales up to support realistic \mbox{integrity-preserving} DNN model training for heavy workloads (deep architectures and millions of training examples) without sustaining a significant performance hit.
To mitigate the time gap between pure TEE (full integrity) and pure GPU (no integrity), we combine random verification of selected computation steps with systematic adjustments of DNN hyperparameters (e.g., a narrow gradient clipping range), hence limiting the attacker's ability to shift the model parameters significantly provided that the step is not selected for verification during its training phase.
Experimental results show the new approach achieves 2X to 20X performance improvement over pure TEE based solution while guaranteeing a very high probability of integrity (e.g., 0.999) with respect to \mbox{state-of-the-art} DNN backdoor attacks.

%% file: 002intro.tex
Every day, Artificial Intelligence (AI) and Deep Learning (DL) are incorporated into some new aspects of the  society. 
As a result, numerous industries increasingly rely on DL models to make decisions, ranging from computer vision to natural language processing~\cite{tesla-autonomous,paper:AIDLCase:Medicine,paper:AIDLCase:5G,paper:AIDLCase:cybersec,paper:AIDLCase:blockchain,paper:AIDLCase:theoretical,paper:AIDLCase:financial,paper:AIDLCase:edgecomputing}.
The training process for these DL models requires a substantial quantity of computational resources (often in a distributed fashion) for training, which traditional CPUs are unable to fulfill. Hence, special hardware, with massive parallel computing capabilities such as GPUs, is often utilized~\cite{cpu-gpu-benchmark}.
At the same time, the DL model building process is increasingly outsourced to the cloud. This is natural, as applying cloud services (e.g., Amazon EC2, Microsoft Azure or Google Cloud) for DL training can be more fiscally palatable for companies by enabling them to focus on the software aspect of their products. 
Nevertheless, such outsourcing raises numerous concerns with respect to the privacy and integrity of the learned models. 
In recognition of the privacy and integrity concerns around DL (and Machine Learning (ML) in general), a considerable amount of research has been dedicated to applied cryptography, in three general areas:
1) \textit{Multi-Party Computation (MPC)} (e.g., \cite{Paper:SecureML}),
2) \textit{Homomorphic Encryption (HE)
} (e.g., \cite{Paper:CryptoNets}),
and 3) \textit{Trusted Execution Environment (TEE)} (e.g., \cite{Paper:Chiron,Paper:Myelin}).
However, the majority of these investigations are limited in that: 1) they are only applicable to simple shallow network models, 2) they are evaluated with datasets that have a small number of records 
(such as MNIST \cite{Dataset:MNIST} and CIFAR10 \cite{ Dataset:CIFAR-10}), and 3) they incur a substantial amount of overhead that is unacceptable for real-life DL training workloads.
In their effort to mitigate some of these problems, and securely move from CPUs to GPUs, \emph{Slalom}~\cite{Paper:Slalom} mainly focus on the computational integrity at the \textit{test} phase while depending on the application context. It can also support enhanced data privacy, however, at a much greater performance cost.

\begin{figure}[t]
\begin{center}
\def\svgwidth{\columnwidth}
\includegraphics[width=\textwidth, 
]
{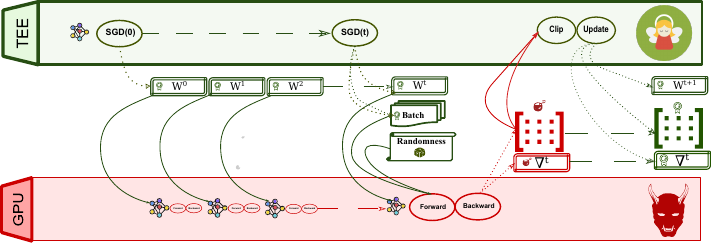}
\caption[]{\label{fig:architecture} 
The main architecture of \textbf{\systemname{}}.
The TEE handles mini-batch selection, layer specific randomness, and parameter initialization. The GPU performs forward and backward passes over the mini-batch and reports the computed gradients to TEE for gradient clipping, weight updates, and preserving intermediate reports in case this step is selected for verification inside the TEE.
}
\end{center}
\end{figure}

To address these limitations, we introduce \textbf{\systemname{}} (See Figure~\ref{fig:architecture}); a framework for 
\emph{integrity-preserving} learning as a service that provides integrity guarantees in outsourced DL model training in TEEs. We assume that only the TEE running in the cloud is trusted, and all the other resources such as GPUs can be controlled by an attacker to launch an attack (e.g., insert a trojan). In this context, our goal is to support the realistic deep learning training workloads while ensuring data and model integrity.
To achieve this goal, we focus on the settings where maintaining the learning process's integrity is critical, while the training data may not contain privacy sensitive information. For example, we may want to build a traffic sign detection model on public traffic sign images and may still like to prevent attacks that can insert trojan during the training phase. 
Furthermore, we want to provide assurances that the model is trained on the specified dataset, with known parameters so that the performance of the model can be replicated and audited for accountability and integrity.

The trivial approach of executing the entire learning process inside a TEE is not scalable since TEEs are much slower compared to GPUs. Furthermore, even the existing performance improvement techniques (e.g., random matrix verification provided in \cite{Paper:Slalom}) are not enough to scale up to large DL model learning settings.

To alleviate the TEE bottleneck, we propose incorporating \emph{random verification} of the computation steps. 
This strategy is based on the observation that it is unnecessary to verify all of the GPU's computation steps. Rather, we only need to verify occasionally to catch any deviation with a very high likelihood.  Given that random verification may itself be insufficient (theoretically, an attacker can launch a successful attack by modifying only a single unconstrained gradient update), we further show how parts of the DL hyperparameter setting process, such as \textit{clipping rate} should be modified to prevent single step attacks, and require a larger number of malicious updates by an attacker that controls the GPU. Simply, \textbf{\systemname{}} limits the amount of change an adversary can inflict on a model through a single SGD step. As a consequence, the adversary is forced to keep attacking while randomly being verified by the TEE. Using the state-of-the-art backdoor attacks, we illustrate that random verification technique can detect attacks with a high probability (e.g., 0.99)  while enabling 2x-20x performance gains compared to pure TEE based solutions.

\begin{description}
\item[The specific contributions of this paper are as follows:]
\end{description}
\begin{itemize}
    \item 
    We introduce the first approach to support integrity-preserving DL training by random verification of stochastic gradient (SGD) steps inside TEE to ensure the integrity of training pipeline data, parameters, computation function, etc. with a high probability.
    \item We illustrate how gradient clipping can be used as a defensive measure against single (or infrequent) step attack in combination with random verification.
    \item We show the effectiveness of our TEE random verification and gradient clipping through extensive experimentation on DNN  backdoor attacks.  
\end{itemize}

%% file: 003back.tex
Our system combines deep learning training on specialized fast hardware such as Graphical Processing Units (GPU) with Intel Software Guard Extensions (SGX) based TEE to ensure the produced model's integrity. 
Details on SGD training and gradient clipping are provided in Appendices~\ref{subsection:training} and~\ref{subsection:gradclipping}.

\subsection{Attacks on DNN Models in Training Phase\label{section:attacks_on_dnn_models_train}}
Attacks on DNN models can be realized during both \textit{training} or \textit{test} phases. 
However, \systemname{} is concerned with integrity/accountability issues during the training phase of DNN models, such that attacks related to testing are out of the scope of this paper since test time attacks (~\cite{paper:adv_ml,paper:adv_ml2}) have been addressed before (e.g., \emph{Slalom}~\cite{Paper:Slalom}).
In the literature, particularly in the computer vision domain, targeted trojan attacks on DNN classification models have become a real concern as deep learning has grown in its adoption. These attacks tend to alter the prediction of models if a specific condition in the input is met. These conditions may be \textit{feature-based}~\cite{paper:backdoor_1,paper:backdoor2,paper:backdoor_clean_label_1} or \textit{instance-based}~\cite{paper:backdoor_3,Poison_Frogs}. 
Recently, trojan attacks have been extended to Reinforcement Learning (RL) and text classification models
~\cite{paper:backdoor_rfl_1,paper:backdoor_nlp_1}.

In practice, these attacks are implemented by manipulating samples during training through data poisoning. For instance,  stamping images with a pattern and modifying its label. Interestingly, these models provide similar competitive classification test accuracy compared to clean models (i.e., models have not been attacked). As a consequence, it is non-trivial to distinguish these trojaned models from non-trojaned ones based on model accuracy alone. To make matters worse, even if the model owner was aware of examples of the trojan trigger pattern, the owner would need to patch the model through re-training to dampen the efficacy of the trojan trigger pattern.
Retraining does not always guarantee complete removal of the trojan behavior from the model.
To date, various techniques have been proposed to diagnose and mitigate of trojaned models. However, all approaches are either based on unrealistic assumptions or are excessively costly. For instance, the Neural Cleanse~\cite{paper:backdoor_mitigation_neural_cleanse} requires access to a sizable sample of clean inputs to reverse-engineer the backdoor and has shown to be successful only for trigger patterns with a relatively small size.
ABS~\cite{paper:backdoor_mitigation_abs}
improves upon Neural Cleanse in that requires a significantly smaller number of samples; however, it assumes that the responsible trojan neurons can activate trojan behavior independently from each other, which is unlikely to be true in practice.

Attacking the training pipeline to inject a trojan(s) in the final model is the cheapest and, thus, is likely the most desirable form of attack for real-world adversaries to launch. As such, throughout this work, we mainly focus on showing our methods' effectiveness in preventing this type of attack from happening.
It should be noted that our method is \textit{orthogonal} to \textit{attack} type and is sufficiently \textit{generic} to catch any continuous attack during the training of a DNN model. \systemname{} relies upon \textit{proactive} training as opposed to \mbox{post-training} or \mbox{deployment-time} methods to assess the health of a DNN model. As we explain later in section~\ref{sec:threat}, we assume that the initial training dataset is provided by an honest user and is free of manipulation. With this as a basis, \systemname{} limits the amount of change an adversary can inflict on a model through a single SGD step. As a consequence, the adversary is forced to keep attacking while randomly being verified by the TEE.
\subsection{
Integrity for DNN Training}
\systemname{}'s main goal is to enable high-integrity training pipeline so that end users are assured that the model is built on the specified dataset, using specified parameters without modification. Thus, the final model users know who built the model, what dataset was used for training, and what algorithms were put in place for building the model.
If, at any point during training, \systemname{} detects a deviation from the specified execution, it will not sign the final model to ascertain its validity. 
\cite{Paper:Slalom} took a first step towards achieving both  \textit{fast} and \textit{reliable} execution in the \textit{test} phase but neglected the training phase. The training phase is far more computationally demanding than the test phase, such that verification of all steps in training requires a substantially longer time. Since parameters keep changing, we cannot benefit from pre-computation. 
 Second, backward pass involves computing gradients for both the inputs and the parameters and takes longer than forward pass. Despite the mentioned hurdles, as our investigation shows, it may not be necessary to verify every step to achieve integrity guarantees with high probability.
\subsection{Intel SGX}
SGX~\cite{paper:intelsgxexplained} is an example of a common TEE that is available in many \mbox{modern-day} computers. As outlined in Table~\ref{tab:symbols},
it provides a secluded hardware reserved area, namely, processor reserved memory (PRM), that is kept private (i.e., it is not readable in plaintext) from the host, or any privileged processes, and is free from \textit{direct} undetected tampering.
It also supports \textit{remote attestation}, such that users can attest the platform and the running code within enclave before provisioning their secrets to a remote server.
Calls from routines that should transition to/from enclave are handled through predefined entry points that are called Ecall/Ocall that must be defined in advance, before building the enclave image. While it provides security and privacy for numerous applications (e.g., ~\cite{paper:enclavedb,paper:sgx_big_matrix,paper:Tensorscone}), due to its limited memory and computational capacity, directly running unmodified applications inside SGX can induce a significant hit on performance. This is especially the case for applications that require large amounts of memory, such as training DNNs.

%% file: 004threat.tex
Attacks on the integrity of DNNs can be orchestrated at different stages of the model learning pipeline (e.g., data collection or training). We assume the TEE node in \systemname{} is trusted, and the bytes stored on the \textit{PRM} are always encrypted and authenticated before they are fetched inside the CPU. 
We assume that the data sent to \systemname{} comes from honest users via a secure/authenticated channel and is devoid of malicious samples.\footnote{Detecting malicious samples is beyond the scope of this work.} For the training phase, we assume that the adversary has complete knowledge about the network structure, learning algorithm, and inputs (after TEE performs an initial pre-processing) to the model.
In our threat model, the adversary is in complete control of the host system's software stack, and hardware (unprotected RAM, GPU), except for the CPU package and its internals. Therefore, the code that runs inside the enclave is free from tampering, and the data that is accessed inside the cache-lines or registers are not accessible to the adversary.
For the inputs supplied to DNN tasks, the adversary is capable of performing insertion, modification, and deletion to influence the final model towards her advantage. 
As a result, an attacker may report wrong gradients as opposed to correctly computed ones.

%% file: 005sys.tex
\systemname{} offers integrity and accountability for the training phase of a DNN model while inducing limited computational overhead. An overview of \systemname{} is illustrated in figure~\ref{fig:architecture}, and we refer the reader to table~\ref{tab:symbols} for symbol descriptions and abbreviations. 

Before the training phase initiates, the training dataset is decrypted and validated inside the TEE. 
Besides, for each SGD step, the randomness regarding the mini-batch selection and the network layers~\cite{paper:droput_layer} are derived within the TEE and supplied to the GPU. As a result, the adversary will face a more constrained environment. Moreover, in our design, the GPU always performs a forward and a backward pass and reports the computed gradients to the TEE. At this point, the TEE clips the gradients and updates the parameters (low overhead operation). Finally, \systemname{} randomly decides to verify the SGD steps within the TEE or not.
\systemname{} is optimized to guarantee a high level of integrity and the correctness of the model while providing an infrastructure that does not suffer from the substantial computational requirements of pure TEE-based solutions.
We assume an honest and authenticated user will send her data encryption key $K_{client}$ (after remote-attestation) to the TEE. Next, the TEE decrypts/verifies the initial encrypted dataset using the $K_{client}$ and supplies the trainer (GPU) the plain-text of the training set. If the TEE fails to detect any violations of the protocol during training, it will sign a message that certifies the final model.(Please see Appendix~\ref{appendix:signature} for more details of the signed message).
Then, during testing and deployment, the user can verify the digital signature of the model.
\begin{description}[style=unboxed,leftmargin=0cm]
\item[Training with \systemname{}]
At the beginning of mini-batch iteration $\mathbf{i}$, TEE supplies the untrusted GPU with the randomness for that iteration.
After completion of the forward and backward passes over the mini-batch, the computed gradients %
are sent to the TEE for clipping and updating the parameters. Next, the TEE integrates the clipped gradients with the parameters of the previous step.
\systemname{} always clips the reported gradients and ensures that they are within a narrow range \textit{so that evolving the model towards the attacker's intended model requires a prolonged malicious intervention by the attacker}.
\systemname{} accepts the reported gradients and only applies the clipped version to the snapshot taken at the specific iteration. If the computation at that step is selected for random verification, then the faulty behavior can be detected. If not, the chance that the model evolved towards the attacker’s desired optima will likely require multiple rounds providing ample opportunity for detection. The verification is done randomly to prevent an attacker from guessing which step is verified.
\item[Probabilistic Verification with \systemname{}]
The TEE  randomly decides whether or not to verify the computation over each mini-batch. If the mini-batch is selected for verification, then the intermediate results are saved, and the verification task is pushed into a verification queue. \textit{Verification by the TEE can take place asynchronously} and it does not halt the computation for future iterations on the GPU.
The authenticity of snapshots is always verified with a key that is derived from a combination of the TEE's session key, $SK_{SGX}^{session}$ and the corresponding iteration.
When the TEE verifies step $\mathbf{i}$, it populates the network parameters with the snapshot it created for the step $\mathbf{i-1}$. It then regenerates the randomness from step $i$ to obtain the batch indices and correctly sets up the per layer randomness. Given that the TEE’s goal is to verify that the reported gradients for step $i$ are correctly computed, \systemname{} \underline{does not} keep track of the activation results. Rather it only requires  the computed gradients, batch mean/std (for BatchNorm layer), and matrix multiplication outcomes (in case random matrix multiplication verification is chosen). These required parameters are saved for verification.

\item[Randomized Matrix Multiplication Verification with \systemname{}]
Matrix Multiplications (MM) take up the bulk of the resource-heavy computations in DNNs. In modern DNN frameworks, convolutional and connected layers computation is implemented in the form of a matrix multiplication in both of the forward and backward passes. Table~\ref{tab:MM-OPS-Verification} in Appendix~\ref{section:MM_OPS_APX} depicts the computations in the forward pass and backward gradient with respect to the weights and previous layers' outputs in the form of a rank 2 tensor multiplication. Fortunately, there exists an efficient verification algorithm for matrix multiplication(~\cite{paper:freivalds}) when the elements of matrices belong to a field. In this work, we leverage these random matrix multiplication verification algorithms as well.
\end{description}

%% file: 006sec.tex
To achieve the integrity goal, we need to derive the probability $p_{v}$ (i.e., the verification probability of each step) to achieve our integrity goal $p_i$ (i.e., the probability that attacker can modify the result without being detected is less than $1-p_{i}$). %
\begin{definition}
Assuming the DNN training requires a total of $B$ steps, for each step report ($R_{b} \; \forall b \; \in \; [1,B] \; \land R_{b} \in \{0,1\}$) has a probability $p_{c}$ for being corrupted ( i.e., $R_{b} = 1$), and the overall integrity requirement probability goal $p_{i}$ (for example $p_{i} = 0.999$).
\end{definition}
\begin{theorem}
\label{theorem:rand_mb_verf}
Given a total of B steps during SGD training, the required probability of choosing a step to verify ($p_{v}$) should be greater than $ B^{-1}(\frac{\log(1 - p_{i})}{\log(1-p_{c})} - 1)$.
\end{theorem}
\begin{remark}
Given each step contains $m$ independent matrix multiplication (MM) operations that is to be repeated $k$ times (independently), and each random entry is chosen from a field of size $|S|$, the probability of error (accepting a wrong MM equality) is less than $ \alpha = \frac{1}{|S|^{mk}}$~\cite{paper:freivalds}.
\end{remark}
\begin{theorem}
\label{theorem:rand_mb_mm_verf}
If random matrix multiplication verification is used, given the configuration of Theorem~\ref{theorem:rand_mb_verf}, the required probability of choosing a step to verify ($p_{v}$) should be greater than $B^{-1}(\frac{\log(1 - p_{i})}{ \log((1+(\alpha-1)p_{c})}-1)$.
\end{theorem}
We refer the reader to Appendix ~\ref{proof:rand_mb_verf} and ~\ref{proof:rand_mb_mm_verf} for complete proofs of the theorems.
The threshold probability in Theorem~\ref{theorem:rand_mb_mm_verf} yields approximately the same values as Theorem~\ref{theorem:rand_mb_verf} when $\alpha \to 0 $. However, the randomized matrix multiplication verification requires a $O(N^2)$ operations (assuming two $N\times N$ matrices)  compared to regular matrix multiplication that requires $O(N^3)$ operations.

%% file: 007eval.tex
\begin{figure}[t]
    \begin{center}
		\begin{tikzpicture}%
		\begin{groupplot}[
		group style={
		    group name=teeperf,
		    group size=4 by 1,
		    /pgf/bar width=9pt,
		    y descriptions at=edge left,
		    horizontal sep=-30pt,
		    },
        ybar=\pgflinewidth,
		ylabel=Throughput (Images/Sec),
		ylabel style={font=\small},
		xticklabels={SGX,SGX\textsuperscript{RMM}},
		xticklabel style={font=\tiny},
		yticklabel style={font=\tiny},
		xtick = data,
		nodes near coords,
		nodes near coords style={font=\small,rotate=90,anchor=west},
		ymax=4.5,
		ytick={.5,1.,1.5,2.,2.5,3.,3.5,4.},
		axis lines*=left,
		width=0.40\textwidth,
		height=.20\textwidth,
		clip=false,
		ytick align=outside,
        legend columns=-1,
        legend entries = {Forward,Backward,Overall},
        legend to name=legendnamed,
        enlarge x limits = {abs=0.8},
		]
		
		\nextgroupplot[xlabel=VGG19,]
		\addplot coordinates {
			(1,0.9338421976)
			(2,1.59760499)
		};
		\addplot coordinates {
			(1,0.4208629359) (2,1.043230545)
		};
		\addplot coordinates {
			(1,0.2901144753) (2,0.6311147749)
		};
		
		\nextgroupplot[xlabel=VGG16,y axis line style={draw opacity=0},ytick style={draw=none},]
		\addplot coordinates {
			(1,1.103218274)
			(2,1.628598474)
		};
		\addplot coordinates {
			(1,0.474668556) (2,1.045704663)
		};
		\addplot coordinates {
			(1,0.3310894556) (2,0.7007663515)
		};
		
		\nextgroupplot[xlabel=ResNet152,y axis line style={draw opacity=0},ytick style={draw=none},]
		\addplot coordinates {
			(1,1.049119427) (2,1.145220243)
		};
		\addplot coordinates {
			(1,0.7253603312) (2,0.9287871736)
		};
		\addplot coordinates {
			(1,0.4286207975) (2,0.5128553853)
		};
		\nextgroupplot[xlabel=ResNet34,y axis line style={draw opacity=0},ytick style={draw=none},]
		\addplot coordinates {
			(1,3.931354617) (2,4.405227973)
		};
		\addplot coordinates {
			(1,2.243294547) (2,3.007872307)
		};
		\addplot coordinates {
			(1,1.428368046) (2,1.787425332)
		};
		
        \end{groupplot}
        \node [above] at (current bounding box.north) {Throughput Performance (ImageNet)};
		\end{tikzpicture}
	    \ref{legendnamed}
	\caption{Throughput of the SGD training step for VGG19,VGG16, ResNet152, and Resnet34 on \mbox{ImageNet} dataset with repect to forward and backward passes. RMM can lead to verification that is twice as fast as full MM verification in case of a VGG architecture.}
	\label{fig:vgg_resnet_imagenet_performance}
	\end{center}
\end{figure}
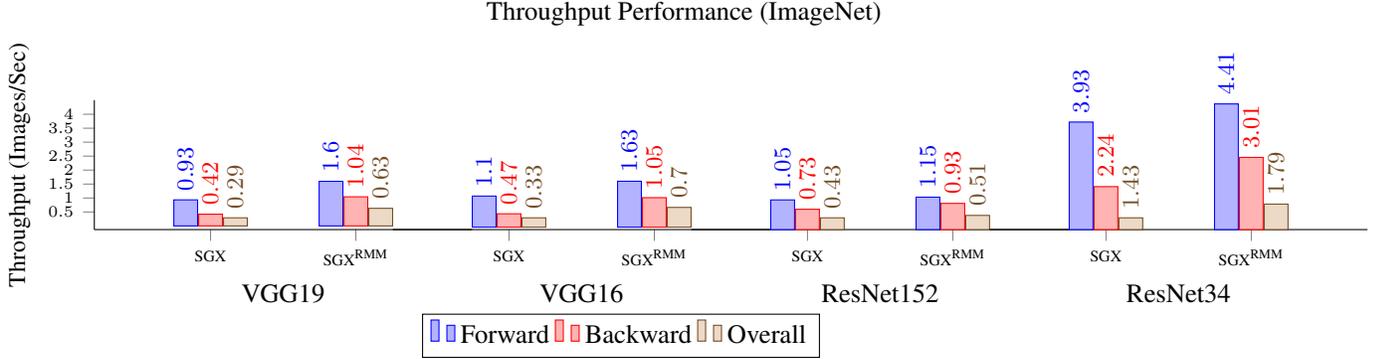
	
\definecolor{color0}{rgb}{0.172549019607843,0.627450980392157,0.172549019607843}
\pgfplotsset{
    boxplots style/.style={
        tick align=outside,
        tick pos=left,
        width=.19\textwidth,
        height=.12\textwidth,
        scale only axis,
        label style ={font=\tiny},
        scaled y ticks = false,
        title style={font=\tiny},
        y tick label style={
        /pgf/number format/.cd,
            fixed,
            fixed zerofill,
            precision=2,
        /tikz/.cd
        },
        clip=false,
        axis x line*=bottom,
        axis y line*=left,
    },
    boxplots box style/.style={
        thick, blue,
    },
    boxplots connectors style/.style={
        thick, black, dash pattern=on 1pt off 3pt on 3pt off 3pt,
    },
    boxplots median style/.style={
        thick, red,
    },
    boxplots outpoints style/.style={
        black, mark=+, mark size=2, mark options={solid,fill opacity=0}, only marks,
    },
    boxplots meanpoints style/.style={
    color0, mark=triangle*, mark size=2, mark options={solid}, only marks,
    },
    boxplots idonknow style/.style={
        black,
    },
}
\begin{table}[t]
\small
\centering
\caption{Trained models with restricted clipping and learning rate}
\begin{tabular}{|c|c|c|c|c|c|}
\hline
\textbf{Dataset} & \textbf{clip} & \textbf{lr} & \textbf{\%clean} & \textbf{\%attack} & \textbf{total}\\
\hline
GTSRB & $10^{-4}$ & $10^{-4}$ & $\geq \%95 $ & $\geq \%85 $ & 124\\
\hline
MNIST & $10^{-4}$ & $5 \times10^{-5}$ & $\geq \%95$ & $\geq \%85$ & 49\\
\hline
CIFAR10 & $5 \times 10^{-4}$ & $10^{-1}$ & $\geq \%90$ & $\geq \%85$ & 94\\
\hline
\end{tabular}
\label{tab:gpu_expers_info}
\end{table}
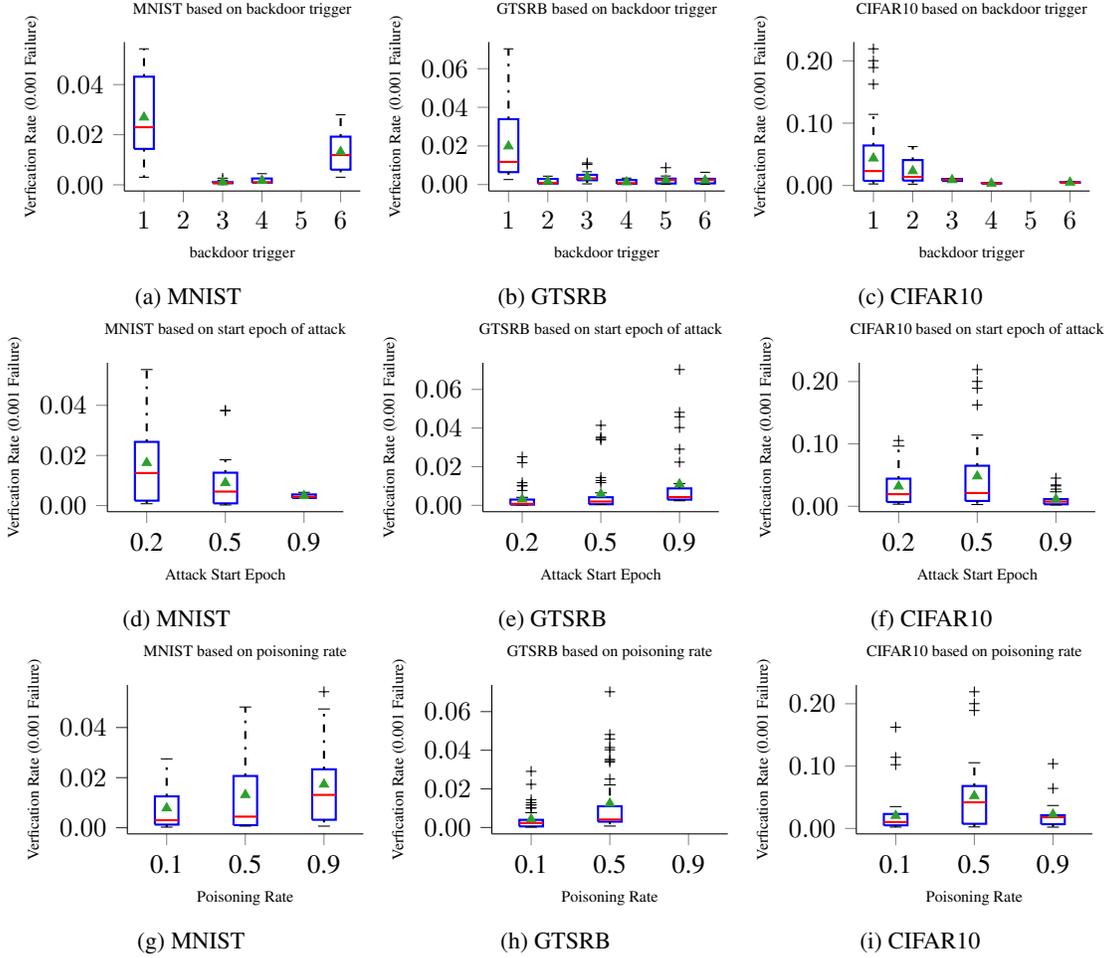
\begin{figure}[t]
    \centering
    \subfloat[MNIST]{%
        \label{fig:bkdr_trig_mnist}
        \input{tex_plots/attack_vs_backdoor_mnist}
    }
    \subfloat[GTSRB]{%
        \label{fig:bkdr_trig_gtsrb}
        \input{tex_plots/attack_vs_backdoor_gtsrb}
    }
    \subfloat[CIFAR10]{%
        \label{fig:bkdr_trig_cifar10}
        \input{tex_plots/attack_vs_backdoor_cifar10}
    }
    
    \subfloat[MNIST]{%
        \label{fig:start_atk_mnist}
        \input{tex_plots/attack_vs_start_mnist}
    }
    \subfloat[GTSRB]{%
        \label{fig:start_atk_gtsrb}
        \input{tex_plots/attack_vs_start_gtsrb}
    }
    \subfloat[CIFAR10]{%
        \label{fig:start_atk_cifar10}
        \input{tex_plots/attack_vs_start_CIFAR10}
    }
    
    \subfloat[MNIST]{%
        \label{fig:poise_rate_mnist}
        \input{tex_plots/attack_vs_pois_mnist}
    }
    \subfloat[GTSRB]{%
        \label{fig:poise_rate_gtsrb}
        \input{tex_plots/attack_vs_pois_gtsrb}
    }
    \subfloat[CIFAR10]{%
        \label{fig:poise_rate_cifar10}
        \input{tex_plots/attack_vs_pois_cifar10}
    }
    \caption[]{A boxplot representation of the verification rates required by TEE for detection failure $\leq 10^{-3}$. Based on the attack \mbox{hyper-parameters}: 1) backdoor trigger type (first row). 2) epoch that attack starts (second row, 20\%$\times$,50\%$\times$, and 90\%$\times|\text{epochs}|)$. 3) backdoor poisoning ratio}
    \label{fig:Dataset_Trigger_Type}
\end{figure}

We executed our experiments on a server with Linux
OS, Intel Xeon CPU E3-1275 v6@3.80GHz, 64GB of RAM and an NVIDIA Quadro P5000 GPU with 16GB of memory. Our attack code is implemented in python 3.6 using the pytorch~\cite{impl:pytorch} library. We use Intel SGX as our platform for TEE. For SGX \mbox{proof-of-concept} implementation, DarkNet~\cite{impl:darknet13} library is significantly modified to run the experiments. Our SGX code has been tested with SDK 2.9 and the code runs inside a docker container in hardware mode. 
Our experiments are designed to investigate two aspects. First, we evaluate the impact of integrating randomized matrix verification and show that it can potentially increase computational efficiency. Second, we analyze the effectiveness of gradient clipping in forcing the attacker to \underline{deviate} from the honest protocol with significant numbers of mini-batch steps. Various attack \mbox{hyper-parameters} (e.g. poisoning rate) were evaluated in determining their importance towards a successful attack with minimal deviation.
This is important because, if the attacker needs to deviate over more mini-batch steps, then the TEE can detect such deviations with a smaller number of random verification steps (i.e., $p_c$ is higher in equation~\ref{eq:rand-sel} ).
\paragraph{TEE Performance}
We tested TEE performance on VGG16, VGG19~(\cite{paper:VGG}),  ResNet152, and ResNet34~(\cite{paper:resnet}) architectures with the  ImageNet~(\cite{Dataset:imagenet_cvpr09}) dataset (see Appendix~\ref{subsec:appdx_tee_cifar10} for results with CIFAR10 dataset(~\cite{Dataset:CIFAR-10})). Figure~\ref{fig:vgg_resnet_imagenet_performance} illustrates the throughput of deep networks. Usually, most of the computation takes place in the convolution layers of the network. However, the backward pass on average yields a smaller throughput since it involves one more MM (weight gradients, and input gradients) than the forward pass which only invokes MM once (i.e., output of convolution or fully-connected layers). The implementation is quite efficient in terms of MM operations which uses both vectorized instructions along with multi-threading. Both VGG networks gives better improvements ($\approx\text{2.2\textbf{X}}$) than ResNet ($\approx\text{1.2}\textbf{X}$). Considering the TEE randomly decides to verify a SGD step with probability $p_{v}$, the overall improvement is approximately multiplied by $\frac{1}{p_{v}}\textbf{X}$. 
For instance, if we assume the attacker only requires to deviate with probability $5.10^{-5}$ (i.e., deviating in five steps out of a hundred thousand of steps), we can detect it with $p_{v}\approx0.1$ (See Figure~\ref{subfig:detect_0.999_ds_1M} in Appendix) while maintaining integrity violation detection probability ($p_{i} > 0.999$). For VGG networks it means approximately $\text{22}\textbf{X}$ improvements compared to a pure TEE-based solution that verifies every step. Nonetheless, as our experiments suggest, we believe that attacking these deep models that are trained on massive datasets should require a significantly higher number of deviations. This in return may result in a much higher performance gain.
\paragraph{Combined Impact of Gradient Clipping and Learning Rate on Attack Success}
As shown in Table~\ref{tab:gpu_expers_info}, we selected models that achieved high performance on both clean and poisoned test samples.
We conducted the attack in the following manner. \footnote{We also experimented with the scenario where an attacker attack at steps chosen randomly. This type of attack was not successful. Hence we do not report those results here.} 
First, the trainer follows the correct protocol (e.g. learning with mini-batch SGD) until $epoch_{attack}$. At this epoch, the attacker starts by injecting a certain number of poisoned samples ($pois_{rate}\times batch\_size$) from every class into the training batch and labels it as the target label. The attacker continues to attack until they achieve a desired threshold in terms of success rate (correctly classifying backdoored samples as the attacker's target label). Once it passes the threshold, the attacker halts the attack and returns to the honest protocol, while observing the decay in attack success rate. If the success rate falls below the desired low-threshold, the attacker transitions back to attack mode and repeats the aforementioned strategy.
CIFAR10~(\cite{Dataset:CIFAR-10}), GTSRB~(\cite{Dataset:GTSRB}) and MNIST~(\cite{Dataset:MNIST}), were used to analyze the impact of multiple factors imposed by the attacker that also aims to maintain an acceptable accuracy over clean test set. For MNIST, and GTSRB, the \textit{Adam}~\cite{paper:adam_optimizer}
optimizer (which requires a smaller initial learning rate), and for the CIFAR10 dataset, \textit{SGD} with momentum are used. Additionally, for MNIST, and GTSRB $10\%$ of the training set was chosen for the validation set to help adjust the learning rate based on the validation set loss. Save for CIFAR10 dataset, the learning rate was set to decay (by tenfold) at fixed epochs $(40,70,100)$.
\paragraph{Backdoor Trigger Pattern}
We applied 6 different backdoor triggers ( all of them are shown in Appendix, Figure~\ref{fig:appdx_CIFAR10_Trigger_Example}). The MNIST dataset only has single channel images, we converted it to a three channel image to apply the triggers 3 to 6.
As shown in Figures ~\ref{fig:Dataset_Trigger_Type}~\cref{fig:bkdr_trig_mnist,fig:bkdr_trig_gtsrb,fig:bkdr_trig_cifar10} the trigger pattern can significantly influence the effectiveness of the attack. The red lines show the median, while the green dots correspond to the mean.
In all of the datasets, the first trigger pattern (Figure~\ref{fig:appdx_CIFAR10_Trigger_Example:trigger1}) was the most effective one. It covers a wider range of pixels compared to the second trigger type (Figure~\ref{fig:appdx_CIFAR10_Trigger_Example:trigger2}). As a consequence, it is more likely for the model to remember the trigger pattern across longer periods of SGD steps. Moreover, because photo filters (e.g. \textit{Instagram}) are popular these days, we investigate the possibility of conducting attacks using some of the filters(or transformations) as the trigger pattern. However, covering a very wide range of pixels does not lead to a stealthy attack, as illustrated for the last four patterns.
These patterns cover the whole input space and transform it to a new one that they share a lot of spacial similarities while only different in tone or scale (e.g. Figure~\ref{fig:appdx_CIFAR10_Trigger_Example:trigger4}). Learning to distinguish inputs that are similar, but only different in their tone is demanding in terms of continuity of the attack. In this case, both of the images (w. and w.o. the trigger) are influencing most of the parameters and filters in a contradictory manner (different classification label). Thus it takes a significant number of steps for the network to learn to distinguish them when gradient clipping is applied.
\paragraph{Attack Start Epoch}
Another major factor influencing the evasiveness of the attack is when attacker starts the attack. For instance, early in the training phase the learning rate will be high, such that a savvy attacker might believe they can avoid low clipping values by initiating their attack. However, if the attack begins too early, then it is unlikely that the model has yet converged. Therefore, beginning the attack too early  may require an unnecessarily high number of (unnecessarily) poisoned batches,
which, in turn, would raise the probability of detection.
Yet even if the attacker was successful, once they halt the attack, the model will likely evolve the parameters back to a clean state relatively quickly, and may require the attacker to re-initiate their attack.
At the same time, given that the system uses a low clipping value, if the attacker waits toward the end of training, the attack is again unlikely to be effective. It would be unlikely that the attack succeeds before the end of the training; particularly due to having a considerably smaller learning rate.
As shown in Figures~\ref{fig:Dataset_Trigger_Type}~\cref{fig:start_atk_mnist,fig:start_atk_gtsrb,fig:start_atk_cifar10},
the best time for the attack is when the model has a relatively low loss on clean training inputs, and the combination of learning rate and clipping value (effective attainable update) could yield the model to move toward attacker’s desired optima. However, for MNIST, which is an easier learning task, attacking early gives the attacker a better chance to launch a stealthier attack. We speculate that this is due to the fast convergence of the model.
After a few epochs, it quickly reaches a stable low training loss value for clean images. As a result, when the attacker concludes the attack (after reaching to the desired threshold), it is generally preserved far longer than the other two datasets. 
\paragraph{Mini-Batch Poisoning Ratio}
As we stated earlier, the $pois_{rate}$ parameter is the ratio of the number of poisoned samples in the batch to the batch size. 
It is one of the critical factors amongst those we investigated. Especially when gradient clipping is used, setting $pois_{rate}$ appropriately can help the attacker by moving more parameters toward the desired optima. However going beyond the ratio $0.9$ (i.e., $pois_{rate} > 0.9$ can impact the training negatively for both clean inputs and poisoned inputs.
As depicted in Figures~\ref{fig:Dataset_Trigger_Type}~\cref{fig:poise_rate_mnist,fig:poise_rate_gtsrb,fig:poise_rate_cifar10}, our experiments suggest that filling more than half the batch with poisoned samples seems to be effective across all datasets. Although for the MNIST dataset, it shows that higher values can slightly perform better, but this is not confirmed by experiments on more complex datasets. For example, for the GTSRB dataset, we could not find a successful attack model with acceptable clean input accuracy.

%% file: tex_plots/attack_vs_backdoor_mnist.tex
\begin{tikzpicture}
    \begin{axis}[
    boxplots style,
    title={MNIST based on backdoor trigger},
    xlabel={backdoor trigger},
    xmin=0.5, xmax=6.5,
    ylabel={Verfication Rate (0.001 Failure)},
    ymin=-0.00245370006023086, ymax=0.056969928853667,
    xtick={1,2,3,4,5,6},
    ]
    \addplot [boxplots box style,]
    table {%
    0.75 0.0143418165580175
    1.25 0.0143418165580175
    1.25 0.0431561713170533
    0.75 0.0431561713170533
    0.75 0.0143418165580175
    };
    \addplot [boxplots connectors style,]
    table {%
    1 0.0143418165580175
    1 0.00302893428949408
    };
    \addplot [boxplots connectors style,]
    table {%
    1 0.0431561713170533
    1 0.0542688548121261
    };
    \addplot [boxplots idonknow style,]
    table {%
    0.875 0.00302893428949408
    1.125 0.00302893428949408
    };
    \addplot [boxplots idonknow style,]
    table {%
    0.875 0.0542688548121261
    1.125 0.0542688548121261
    };
    \addplot [boxplots box style,]
    table {%
    1.75 nan
    2.25 nan
    2.25 nan
    1.75 nan
    1.75 nan
    };
    \addplot [boxplots connectors style,]
    table {%
    2 nan
    2 nan
    };
    \addplot [boxplots connectors style,]
    table {%
    2 nan
    2 nan
    };
    \addplot [boxplots idonknow style,]
    table {%
    1.875 nan
    2.125 nan
    };
    \addplot [boxplots idonknow style,]
    table {%
    1.875 nan
    2.125 nan
    };
    \addplot [boxplots box style,]
    table {%
    2.75 0.000669652548737133
    3.25 0.000669652548737133
    3.25 0.0010661286926804
    2.75 0.0010661286926804
    2.75 0.000669652548737133
    };
    \addplot [boxplots connectors style,]
    table {%
    3 0.000669652548737133
    3 0.000247373981309948
    };
    \addplot [boxplots connectors style,]
    table {%
    3 0.0010661286926804
    3 0.0010661286926804
    };
    \addplot [boxplots idonknow style,]
    table {%
    2.875 0.000247373981309948
    3.125 0.000247373981309948
    };
    \addplot [boxplots idonknow style,]
    table {%
    2.875 0.0010661286926804
    3.125 0.0010661286926804
    };
    \addplot [boxplots outpoints style,]
    table {%
    3 0.00183820025325233
    3 0.00265282361390082
    };
    \addplot [boxplots box style,]
    table {%
    3.75 0.000849988967028809
    4.25 0.000849988967028809
    4.25 0.00252612749458899
    3.75 0.00252612749458899
    3.75 0.000849988967028809
    };
    \addplot [boxplots connectors style,]
    table {%
    4 0.000849988967028809
    4 0.000608518935961478
    };
    \addplot [boxplots connectors style,]
    table {%
    4 0.00252612749458899
    4 0.00444768782548278
    };
    \addplot [boxplots idonknow style,]
    table {%
    3.875 0.000608518935961478
    4.125 0.000608518935961478
    };
    \addplot [boxplots idonknow style,]
    table {%
    3.875 0.00444768782548278
    4.125 0.00444768782548278
    };
    \addplot [boxplots box style,]
    table {%
    4.75 nan
    5.25 nan
    5.25 nan
    4.75 nan
    4.75 nan
    };
    \addplot [boxplots connectors style,]
    table {%
    5 nan
    5 nan
    };
    \addplot [boxplots connectors style,]
    table {%
    5 nan
    5 nan
    };
    \addplot [boxplots idonknow style,]
    table {%
    4.875 nan
    5.125 nan
    };
    \addplot [boxplots idonknow style,]
    table {%
    4.875 nan
    5.125 nan
    };
    \addplot [boxplots box style,]
    table {%
    5.75 0.00604985521363835
    6.25 0.00604985521363835
    6.25 0.0192248954290242
    5.75 0.0192248954290242
    5.75 0.00604985521363835
    };
    \addplot [boxplots connectors style,]
    table {%
    6 0.00604985521363835
    6 0.00302893428949408
    };
    \addplot [boxplots connectors style,]
    table {%
    6 0.0192248954290242
    6 0.0279481950492084
    };
    \addplot [boxplots idonknow style,]
    table {%
    5.875 0.00302893428949408
    6.125 0.00302893428949408
    };
    \addplot [boxplots idonknow style,]
    table {%
    5.875 0.0279481950492084
    6.125 0.0279481950492084
    };
    \addplot [boxplots median style,]
    table {%
    0.75 0.0229977695899067
    1.25 0.0229977695899067
    };
    \addplot [boxplots meanpoints style,]
    table {%
    1 0.0269205783531176
    };
    \addplot [boxplots median style,]
    table {%
    1.75 nan
    2.25 nan
    };
    \addplot [boxplots meanpoints style,]
    table {%
    2 nan
    };
    \addplot [boxplots median style,]
    table {%
    2.75 0.000734734916010328
    3.25 0.000734734916010328
    };
    \addplot [boxplots meanpoints style,]
    table {%
    3 0.00104167420669197
    };
    \addplot [boxplots median style,]
    table {%
    3.75 0.00102976719569372
    4.25 0.00102976719569372
    };
    \addplot [boxplots meanpoints style,]
    table {%
    4 0.00169769095969482
    };
    \addplot [boxplots median style,]
    table {%
    4.75 nan
    5.25 nan
    };
    \addplot [boxplots meanpoints style,]
    table {%
    5 nan
    };
    \addplot [boxplots median style,]
    table {%
    5.75 0.0118924685808583
    6.25 0.0118924685808583
    };
    \addplot [boxplots meanpoints style,]
    table {%
    6 0.0131687781473573
    };
\end{axis}
\end{tikzpicture}

%% file: tex_plots/attack_vs_backdoor_gtsrb.tex
\begin{tikzpicture}
    \begin{axis}[
    boxplots style,
    title={GTSRB based on backdoor trigger},
    xlabel={backdoor trigger},
    xmin=0.5, xmax=6.5,
    ylabel={Verfication Rate (0.001 Failure)},
    ymin=-0.00332509731804029, ymax=0.0737830620080433,
    xtick={1,2,3,4,5,6},
    ]
    \addplot [boxplots box style,]
    table {%
    0.75 0.00643968438767685
    1.25 0.00643968438767685
    1.25 0.0338713999390099
    0.75 0.0338713999390099
    0.75 0.00643968438767685
    };
    \addplot [boxplots connectors style,]
    table {%
    1 0.00643968438767685
    1 0.00260380894154466
    };
    \addplot [boxplots connectors style,]
    table {%
    1 0.0338713999390099
    1 0.0702781456750395
    };
    \addplot [boxplots idonknow style,]
    table {%
    0.875 0.00260380894154466
    1.125 0.00260380894154466
    };
    \addplot [boxplots idonknow style,]
    table {%
    0.875 0.0702781456750395
    1.125 0.0702781456750395
    };
    \addplot [boxplots box style,]
    table {%
    1.75 0.000518827148304547
    2.25 0.000518827148304547
    2.25 0.00292026252791148
    1.75 0.00292026252791148
    1.75 0.000518827148304547
    };
    \addplot [boxplots connectors style,]
    table {%
    2 0.000518827148304547
    2 0.000179819014963515
    };
    \addplot [boxplots connectors style,]
    table {%
    2 0.00292026252791148
    2 0.00433826340297488
    };
    \addplot [boxplots idonknow style,]
    table {%
    1.875 0.000179819014963515
    2.125 0.000179819014963515
    };
    \addplot [boxplots idonknow style,]
    table {%
    1.875 0.00433826340297488
    2.125 0.00433826340297488
    };
    \addplot [boxplots box style,]
    table {%
    2.75 0.00223204060726197
    3.25 0.00223204060726197
    3.25 0.00503968771598192
    2.75 0.00503968771598192
    2.75 0.00223204060726197
    };
    \addplot [boxplots connectors style,]
    table {%
    3 0.00223204060726197
    3 0.000365465047696329
    };
    \addplot [boxplots connectors style,]
    table {%
    3 0.00503968771598192
    3 0.00665542346174198
    };
    \addplot [boxplots idonknow style,]
    table {%
    2.875 0.000365465047696329
    3.125 0.000365465047696329
    };
    \addplot [boxplots idonknow style,]
    table {%
    2.875 0.00665542346174198
    3.125 0.00665542346174198
    };
    \addplot [boxplots outpoints style,]
    table {%
    3 0.0111886894467033
    3 0.0104534986289164
    };
    \addplot [boxplots box style,]
    table {%
    3.75 0.000421012873553643
    4.25 0.000421012873553643
    4.25 0.00234715610604458
    3.75 0.00234715610604458
    3.75 0.000421012873553643
    };
    \addplot [boxplots connectors style,]
    table {%
    4 0.000421012873553643
    4 0.000179819014963515
    };
    \addplot [boxplots connectors style,]
    table {%
    4 0.00234715610604458
    4 0.00333276414394966
    };
    \addplot [boxplots idonknow style,]
    table {%
    3.875 0.000179819014963515
    4.125 0.000179819014963515
    };
    \addplot [boxplots idonknow style,]
    table {%
    3.875 0.00333276414394966
    4.125 0.00333276414394966
    };
    \addplot [boxplots box style,]
    table {%
    4.75 0.000548313961815858
    5.25 0.000548313961815858
    5.25 0.00308044432676493
    4.75 0.00308044432676493
    4.75 0.000548313961815858
    };
    \addplot [boxplots connectors style,]
    table {%
    5 0.000548313961815858
    5 0.000179819014963515
    };
    \addplot [boxplots connectors style,]
    table {%
    5 0.00308044432676493
    5 0.00442522001736041
    };
    \addplot [boxplots idonknow style,]
    table {%
    4.875 0.000179819014963515
    5.125 0.000179819014963515
    };
    \addplot [boxplots idonknow style,]
    table {%
    4.875 0.00442522001736041
    5.125 0.00442522001736041
    };
    \addplot [boxplots outpoints style,]
    table {%
    5 0.00873725797608692
    };
    \addplot [boxplots box style,]
    table {%
    5.75 0.00059963954532129
    6.25 0.00059963954532129
    6.25 0.00297156737099151
    5.75 0.00297156737099151
    5.75 0.00059963954532129
    };
    \addplot [boxplots connectors style,]
    table {%
    6 0.00059963954532129
    6 0.000179819014963515
    };
    \addplot [boxplots connectors style,]
    table {%
    6 0.00297156737099151
    6 0.00634260721081539
    };
    \addplot [boxplots idonknow style,]
    table {%
    5.875 0.000179819014963515
    6.125 0.000179819014963515
    };
    \addplot [boxplots idonknow style,]
    table {%
    5.875 0.00634260721081539
    6.125 0.00634260721081539
    };
    \addplot [boxplots median style,]
    table {%
    0.75 0.0117409482190389
    1.25 0.0117409482190389
    };
    \addplot [boxplots meanpoints style,]
    table {%
    1 0.0198556715625214
    };
    \addplot [boxplots median style,]
    table {%
    1.75 0.000891096873319013
    2.25 0.000891096873319013
    };
    \addplot [boxplots meanpoints style,]
    table {%
    2 0.00167069276018917
    };
    \addplot [boxplots median style,]
    table {%
    2.75 0.00322215007837473
    3.25 0.00322215007837473
    };
    \addplot [boxplots meanpoints style,]
    table {%
    3 0.00391120478586634
    };
    \addplot [boxplots median style,]
    table {%
    3.75 0.00074495506998012
    4.25 0.00074495506998012
    };
    \addplot [boxplots meanpoints style,]
    table {%
    4 0.00124329348791212
    };
    \addplot [boxplots median style,]
    table {%
    4.75 0.00234715610604458
    5.25 0.00234715610604458
    };
    \addplot [boxplots meanpoints style,]
    table {%
    5 0.00227050190229738
    };
    \addplot [boxplots median style,]
    table {%
    5.75 0.00234352079423363
    6.25 0.00234352079423363
    };
    \addplot [boxplots meanpoints style,]
    table {%
    6 0.00219271800183884
    };
    \end{axis}
\end{tikzpicture}

%% file: tex_plots/attack_vs_backdoor_cifar10.tex
\begin{tikzpicture}
    \begin{axis}[
    boxplots style,
    title={CIFAR10 based on backdoor trigger},
    xlabel={backdoor trigger},
    xtick={1,2,3,4,5,6},
    xmin=0.5, xmax=6.5,
    ylabel={Verfication Rate (0.001 Failure)},
    ymin=-0.00879455502637059, ymax=0.230077845690619,
    ]
    \addplot [boxplots box style,]
    table {%
    0.75 0.00744255512798038
    1.25 0.00744255512798038
    1.25 0.0641843595514223
    0.75 0.0641843595514223
    0.75 0.00744255512798038
    };
    \addplot [boxplots connectors style,]
    table {%
    1 0.00744255512798038
    1 0.00245114604631773
    };
    \addplot [boxplots connectors style,]
    table {%
    1 0.0641843595514223
    1 0.11410395649095
    };
    \addplot [boxplots idonknow style,]
    table {%
    0.875 0.00245114604631773
    1.125 0.00245114604631773
    };
    \addplot [boxplots idonknow style,]
    table {%
    0.875 0.11410395649095
    1.125 0.11410395649095
    };
    \addplot [boxplots outpoints style,]
    table {%
    1 0.162461606349661
    1 0.219220009294392
    1 0.200150980820134
    1 0.189179758886118
    };
    \addplot [boxplots box style,]
    table {%
    1.75 0.00789790230788088
    2.25 0.00789790230788088
    2.25 0.0408198431378555
    1.75 0.0408198431378555
    1.75 0.00789790230788088
    };
    \addplot [boxplots connectors style,]
    table {%
    2 0.00789790230788088
    2 0.00206328136985622
    };
    \addplot [boxplots connectors style,]
    table {%
    2 0.0408198431378555
    2 0.0624397925536437
    };
    \addplot [boxplots idonknow style,]
    table {%
    1.875 0.00206328136985622
    2.125 0.00206328136985622
    };
    \addplot [boxplots idonknow style,]
    table {%
    1.875 0.0624397925536437
    2.125 0.0624397925536437
    };
    \addplot [boxplots box style,]
    table {%
    2.75 0.00731154537357014
    3.25 0.00731154537357014
    3.25 0.0104851409072536
    2.75 0.0104851409072536
    2.75 0.00731154537357014
    };
    \addplot [boxplots connectors style,]
    table {%
    3 0.00731154537357014
    3 0.00641511795694504
    };
    \addplot [boxplots connectors style,]
    table {%
    3 0.0104851409072536
    3 0.0109168925249784
    };
    \addplot [boxplots idonknow style,]
    table {%
    2.875 0.00641511795694504
    3.125 0.00641511795694504
    };
    \addplot [boxplots idonknow style,]
    table {%
    2.875 0.0109168925249784
    3.125 0.0109168925249784
    };
    \addplot [boxplots box style,]
    table {%
    3.75 0.00317169436615547
    4.25 0.00317169436615547
    4.25 0.00348210084223934
    3.75 0.00348210084223934
    3.75 0.00317169436615547
    };
    \addplot [boxplots connectors style,]
    table {%
    4 0.00317169436615547
    4 0.00272389417822854
    };
    \addplot [boxplots connectors style,]
    table {%
    4 0.00348210084223934
    4 0.00391519855055994
    };
    \addplot [boxplots idonknow style,]
    table {%
    3.875 0.00272389417822854
    4.125 0.00272389417822854
    };
    \addplot [boxplots idonknow style,]
    table {%
    3.875 0.00391519855055994
    4.125 0.00391519855055994
    };
    \addplot [boxplots box style,]
    table {%
    4.75 nan
    5.25 nan
    5.25 nan
    4.75 nan
    4.75 nan
    };
    \addplot [boxplots connectors style,]
    table {%
    5 nan
    5 nan
    };
    \addplot [boxplots connectors style,]
    table {%
    5 nan
    5 nan
    };
    \addplot [boxplots idonknow style,]
    table {%
    4.875 nan
    5.125 nan
    };
    \addplot [boxplots idonknow style,]
    table {%
    4.875 nan
    5.125 nan
    };
    \addplot [boxplots box style,]
    table {%
    5.75 0.00429086631517464
    6.25 0.00429086631517464
    6.25 0.0052828119341779
    5.75 0.0052828119341779
    5.75 0.00429086631517464
    };
    \addplot [boxplots connectors style,]
    table {%
    6 0.00429086631517464
    6 0.00424717009896121
    };
    \addplot [boxplots connectors style,]
    table {%
    6 0.0052828119341779
    6 0.006642265469992
    };
    \addplot [boxplots idonknow style,]
    table {%
    5.875 0.00424717009896121
    6.125 0.00424717009896121
    };
    \addplot [boxplots idonknow style,]
    table {%
    5.875 0.006642265469992
    6.125 0.006642265469992
    };
    \addplot [boxplots median style,]
    table {%
    0.75 0.0231845561712766
    1.25 0.0231845561712766
    };
    \addplot [boxplots meanpoints style,]
    table {%
    1 0.043640429123753
    };
    \addplot [boxplots median style,]
    table {%
    1.75 0.0138127105156721
    2.25 0.0138127105156721
    };
    \addplot [boxplots meanpoints style,]
    table {%
    2 0.0232105629392626
    };
    \addplot [boxplots median style,]
    table {%
    2.75 0.00923380660701271
    3.25 0.00923380660701271
    };
    \addplot [boxplots meanpoints style,]
    table {%
    3 0.00889544461095088
    };
    \addplot [boxplots median style,]
    table {%
    3.75 0.00332934801746512
    4.25 0.00332934801746512
    };
    \addplot [boxplots meanpoints style,]
    table {%
    4 0.00332444719092968
    };
    \addplot [boxplots median style,]
    table {%
    4.75 nan
    5.25 nan
    };
    \addplot [boxplots meanpoints style,]
    table {%
    5 nan
    };
    \addplot [boxplots median style,]
    table {%
    5.75 0.00462655893618754
    6.25 0.00462655893618754
    };
    \addplot [boxplots meanpoints style,]
    table {%
    6 0.00495190728626369
    };
\end{axis}
\end{tikzpicture}

%% file: tex_plots/attack_vs_start_mnist.tex
\begin{tikzpicture}
    \begin{axis}[
    boxplots style,
    title={MNIST based on start epoch of attack},
    xlabel={Attack Start Epoch},
    xmin=0.5, xmax=3.5,
    xtick={1,2,3},
    xticklabels={0.2,0.5,0.9},
    ylabel={Verfication Rate (0.001 Failure)},
    ymin=-0.00245370006023086, ymax=0.056969928853667,
    ]
    \addplot [boxplots box style,]
    table {%
    0.85 0.00195538516699673
    1.15 0.00195538516699673
    1.15 0.0254121709720141
    0.85 0.0254121709720141
    0.85 0.00195538516699673
    };
    \addplot [boxplots connectors style,]
    table {%
    1 0.00195538516699673
    1 0.000731690000170859
    };
    \addplot [boxplots connectors style,]
    table {%
    1 0.0254121709720141
    1 0.0542688548121261
    };
    \addplot [boxplots idonknow style,]
    table {%
    0.925 0.000731690000170859
    1.075 0.000731690000170859
    };
    \addplot [boxplots idonknow style,]
    table {%
    0.925 0.0542688548121261
    1.075 0.0542688548121261
    };
    \addplot [boxplots box style,]
    table {%
    1.85 0.000883913719895573
    2.15 0.000883913719895573
    2.15 0.0131136878610876
    1.85 0.0131136878610876
    1.85 0.000883913719895573
    };
    \addplot [boxplots connectors style,]
    table {%
    2 0.000883913719895573
    2 0.000247373981309948
    };
    \addplot [boxplots connectors style,]
    table {%
    2 0.0131136878610876
    2 0.0182723444950646
    };
    \addplot [boxplots idonknow style,]
    table {%
    1.925 0.000247373981309948
    2.075 0.000247373981309948
    };
    \addplot [boxplots idonknow style,]
    table {%
    1.925 0.0182723444950646
    2.075 0.0182723444950646
    };
    \addplot [boxplots outpoints style,]
    table {%
    2 0.0379178591234929
    2 0.0379178591234929
    };
    \addplot [boxplots box style,]
    table {%
    2.85 0.00302893428949408
    3.15 0.00302893428949408
    3.15 0.0044390365950888
    2.85 0.0044390365950888
    2.85 0.00302893428949408
    };
    \addplot [boxplots connectors style,]
    table {%
    3 0.00302893428949408
    3 0.00302893428949408
    };
    \addplot [boxplots connectors style,]
    table {%
    3 0.0044390365950888
    3 0.00528080664359367
    };
    \addplot [boxplots idonknow style,]
    table {%
    2.925 0.00302893428949408
    3.075 0.00302893428949408
    };
    \addplot [boxplots idonknow style,]
    table {%
    2.925 0.00528080664359367
    3.075 0.00528080664359367
    };
    \addplot [boxplots median style,]
    table {%
    0.85 0.0129844813637528
    1.15 0.0129844813637528
    };
    \addplot [boxplots meanpoints style,]
    table {%
    1 0.017065695443086
    };
    \addplot [boxplots median style,]
    table {%
    1.85 0.00558044697998636
    2.15 0.00558044697998636
    };
    \addplot [boxplots meanpoints style,]
    table {%
    2 0.00908001517251028
    };
    \addplot [boxplots median style,]
    table {%
    2.85 0.00350041193951295
    3.15 0.00350041193951295
    };
    \addplot [boxplots meanpoints style,]
    table {%
    3 0.00385562475143672
    };
    \end{axis}
\end{tikzpicture}

%% file: tex_plots/attack_vs_start_gtsrb.tex
\begin{tikzpicture}
    \begin{axis}[
    boxplots style,
    title={GTSRB based on start epoch of attack},
    xlabel={Attack Start Epoch},
    xmin=0.5, xmax=3.5,
    xtick={1,2,3},
    xticklabels={0.2,0.5,0.9},
    y grid style={white!80!black},
    ylabel={Verfication Rate (0.001 Failure)},
    ymin=-0.00332509731804029, ymax=0.0737830620080433,
    ]
    \addplot [boxplots box style,]
    table {%
    0.85 0.000332406281640599
    1.15 0.000332406281640599
    1.15 0.00297076527931703
    0.85 0.00297076527931703
    0.85 0.000332406281640599
    };
    \addplot [boxplots connectors style,]
    table {%
    1 0.000332406281640599
    1 0.000179819014963515
    };
    \addplot [boxplots connectors style,]
    table {%
    1 0.00297076527931703
    1 0.00412384408724811
    };
    \addplot [boxplots idonknow style,]
    table {%
    0.925 0.000179819014963515
    1.075 0.000179819014963515
    };
    \addplot [boxplots idonknow style,]
    table {%
    0.925 0.00412384408724811
    1.075 0.00412384408724811
    };
    \addplot [boxplots outpoints style,]
    table {%
    1 0.0102400947079447
    1 0.0076652750016832
    1 0.0116792318570615
    1 0.0100158977692191
    1 0.0250936069359874
    1 0.0220543510827176
    };
    \addplot [boxplots box style,]
    table {%
    1.85 0.000548313961815858
    2.15 0.000548313961815858
    2.15 0.00418584661084135
    1.85 0.00418584661084135
    1.85 0.000548313961815858
    };
    \addplot [boxplots connectors style,]
    table {%
    2 0.000548313961815858
    2 0.000421012873553643
    };
    \addplot [boxplots connectors style,]
    table {%
    2 0.00418584661084135
    2 0.00643968438767685
    };
    \addplot [boxplots idonknow style,]
    table {%
    1.925 0.000421012873553643
    2.075 0.000421012873553643
    };
    \addplot [boxplots idonknow style,]
    table {%
    1.925 0.00643968438767685
    2.075 0.00643968438767685
    };
    \addplot [boxplots outpoints style,]
    table {%
    2 0.0117409482190389
    2 0.0128516088624823
    2 0.0143853641377161
    2 0.0352674444237295
    2 0.0346782872574729
    2 0.0338713999390099
    2 0.0414036861073098
    };
    \addplot [boxplots box style,]
    table {%
    2.85 0.00292026252791148
    3.15 0.00292026252791148
    3.15 0.00875478322772616
    2.85 0.00875478322772616
    2.85 0.00292026252791148
    };
    \addplot [boxplots connectors style,]
    table {%
    3 0.00292026252791148
    3 0.00234715610604458
    };
    \addplot [boxplots connectors style,]
    table {%
    3 0.00875478322772616
    3 0.0111886894467033
    };
    \addplot [boxplots idonknow style,]
    table {%
    2.925 0.00234715610604458
    3.075 0.00234715610604458
    };
    \addplot [boxplots idonknow style,]
    table {%
    2.925 0.0111886894467033
    3.075 0.0111886894467033
    };
    \addplot [boxplots outpoints style,]
    table {%
    3 0.0290607027003359
    3 0.0223437556526394
    3 0.0458424383917771
    3 0.0402393167582724
    3 0.0481491014475849
    3 0.0702781456750395
    };
    \addplot [boxplots median style,]
    table {%
    0.85 0.000749285682908852
    1.15 0.000749285682908852
    };
    \addplot [boxplots meanpoints style,]
    table {%
    1 0.00310932693562178
    };
    \addplot [boxplots median style,]
    table {%
    1.85 0.00195423343580383
    2.15 0.00195423343580383
    };
    \addplot [boxplots meanpoints style,]
    table {%
    2 0.00584762897420138
    };
    \addplot [boxplots median style,]
    table {%
    2.85 0.00426284336430836
    3.15 0.00426284336430836
    };
    \addplot [boxplots meanpoints style,]
    table {%
    3 0.0109189129471635
    };
    \end{axis}
\end{tikzpicture}

%% file: tex_plots/attack_vs_start_CIFAR10.tex
\begin{tikzpicture}
    \begin{axis}[
        boxplots style,
        title={CIFAR10 based on start epoch of attack},
        xlabel={Attack Start Epoch},
        xmin=0.5, xmax=3.5,
        xtick={1,2,3},
        xticklabels={0.2,0.5,0.9},
        y grid style={white!80!black},
        ylabel={Verfication Rate (0.001 Failure)},
        ymin=-0.00879455502637059, ymax=0.230077845690619,
    ]
    \addplot [boxplots box style,]
    table {%
    0.85 0.00676384352199727
    1.15 0.00676384352199727
    1.15 0.0442185802998975
    0.85 0.0442185802998975
    0.85 0.00676384352199727
    };
    \addplot [boxplots connectors style,]
    table {%
    1 0.00676384352199727
    1 0.00332096109546444
    };
    \addplot [boxplots connectors style,]
    table {%
    1 0.0442185802998975
    1 0.0965381423333022
    };
    \addplot [boxplots idonknow style,]
    table {%
    0.925 0.00332096109546444
    1.075 0.00332096109546444
    };
    \addplot [boxplots idonknow style,]
    table {%
    0.925 0.0965381423333022
    1.075 0.0965381423333022
    };
    \addplot [boxplots outpoints style,]
    table {%
    1 0.105388094624962
    };
    \addplot [boxplots box style,]
    table {%
    1.85 0.00846189814942133
    2.15 0.00846189814942133
    2.15 0.0649493383169096
    1.85 0.0649493383169096
    1.85 0.00846189814942133
    };
    \addplot [boxplots connectors style,]
    table {%
    2 0.00846189814942133
    2 0.00272389417822854
    };
    \addplot [boxplots connectors style,]
    table {%
    2 0.0649493383169096
    2 0.11410395649095
    };
    \addplot [boxplots idonknow style,]
    table {%
    1.925 0.00272389417822854
    2.075 0.00272389417822854
    };
    \addplot [boxplots idonknow style,]
    table {%
    1.925 0.11410395649095
    2.075 0.11410395649095
    };
    \addplot [boxplots outpoints style,]
    table {%
    2 0.162461606349661
    2 0.219220009294392
    2 0.200150980820134
    2 0.189179758886118
    };
    \addplot [boxplots box style,]
    table {%
    2.85 0.00314249579341522
    3.15 0.00314249579341522
    3.15 0.0114212461545736
    2.85 0.0114212461545736
    2.85 0.00314249579341522
    };
    \addplot [boxplots connectors style,]
    table {%
    3 0.00314249579341522
    3 0.00206328136985622
    };
    \addplot [boxplots connectors style,]
    table {%
    3 0.0114212461545736
    3 0.0231845561712766
    };
    \addplot [boxplots idonknow style,]
    table {%
    2.925 0.00206328136985622
    3.075 0.00206328136985622
    };
    \addplot [boxplots idonknow style,]
    table {%
    2.925 0.0231845561712766
    3.075 0.0231845561712766
    };
    \addplot [boxplots outpoints style,]
    table {%
    3 0.0326643202922751
    3 0.0450748862315014
    3 0.0274470704160263
    3 0.0341229543023181
    };
    \addplot [boxplots median style,]
    table {%
    0.85 0.0193572032366153
    1.15 0.0193572032366153
    };
    \addplot [boxplots meanpoints style,]
    table {%
    1 0.0319212373580055
    };
    \addplot [boxplots median style,]
    table {%
    1.85 0.021213495057815
    2.15 0.021213495057815
    };
    \addplot [boxplots meanpoints style,]
    table {%
    2 0.0481316716627498
    };
    \addplot [boxplots median style,]
    table {%
    2.85 0.00744255512798038
    3.15 0.00744255512798038
    };
    \addplot [boxplots meanpoints style,,]
    table {%
    3 0.0114058752314802
    };
    \end{axis}
\end{tikzpicture}

%% file: tex_plots/attack_vs_pois_mnist.tex
\begin{tikzpicture}
    \begin{axis}[
    boxplots style,
    title={MNIST based on poisoning rate},
    xlabel={Poisoning Rate},
    xmin=0.5, xmax=3.5,
    xtick={1,2,3},
    xticklabels={0.1,0.5,0.9},
    ylabel={Verfication Rate (0.001 Failure)},
    ymin=-0.00245370006023086, ymax=0.056969928853667,
    ytick style={color=white!15!black}
    ]
    \addplot [boxplots box style,]
    table {%
    0.85 0.00125914658282338
    1.15 0.00125914658282338
    1.15 0.0125308532090776
    0.85 0.0125308532090776
    0.85 0.00125914658282338
    };
    \addplot [boxplots connectors style,]
    table {%
    1 0.00125914658282338
    1 0.000247373981309948
    };
    \addplot [boxplots connectors style,]
    table {%
    1 0.0125308532090776
    1 0.0274952527253446
    };
    \addplot [boxplots idonknow style,]
    table {%
    0.925 0.000247373981309948
    1.075 0.000247373981309948
    };
    \addplot [boxplots idonknow style,]
    table {%
    0.925 0.0274952527253446
    1.075 0.0274952527253446
    };
    \addplot [boxplots box style,]
    table {%
    1.85 0.00104584505024181
    2.15 0.00104584505024181
    2.15 0.020664862455409
    1.85 0.020664862455409
    1.85 0.00104584505024181
    };
    \addplot [boxplots connectors style,]
    table {%
    2 0.00104584505024181
    2 0.000669652548737133
    };
    \addplot [boxplots connectors style,]
    table {%
    2 0.020664862455409
    2 0.048183050229769
    };
    \addplot [boxplots idonknow style,]
    table {%
    1.925 0.000669652548737133
    2.075 0.000669652548737133
    };
    \addplot [boxplots idonknow style,]
    table {%
    1.925 0.048183050229769
    2.075 0.048183050229769
    };
    \addplot [boxplots box style,]
    table {%
    2.85 0.00319185141244126
    3.15 0.00319185141244126
    3.15 0.0233290892186836
    2.85 0.0233290892186836
    2.85 0.00319185141244126
    };
    \addplot [boxplots connectors style,]
    table {%
    3 0.00319185141244126
    3 0.000649546632426933
    };
    \addplot [boxplots connectors style,]
    table {%
    3 0.0233290892186836
    3 0.047393263259686
    };
    \addplot [boxplots idonknow style,]
    table {%
    2.925 0.000649546632426933
    3.075 0.000649546632426933
    };
    \addplot [boxplots idonknow style,]
    table {%
    2.925 0.047393263259686
    3.075 0.047393263259686
    };
    \addplot [boxplots outpoints style,]
    table {%
    3 0.0542688548121261
    };
    \addplot [boxplots median style,]
    table {%
    0.85 0.00302893428949408
    1.15 0.00302893428949408
    };
    \addplot [boxplots meanpoints style,]
    table {%
    1 0.00783431417423395
    };
    \addplot [boxplots median style,]
    table {%
    1.85 0.00444336221028579
    2.15 0.00444336221028579
    };
    \addplot [boxplots meanpoints style,]
    table {%
    2 0.0131255494932007
    };
    \addplot [boxplots median style,]
    table {%
    2.85 0.0131136878610876
    3.15 0.0131136878610876
    };
    \addplot [boxplots meanpoints style,]
    table {%
    3 0.0173224531966964
    };
\end{axis}
\end{tikzpicture}

%% file: tex_plots/attack_vs_pois_gtsrb.tex
\begin{tikzpicture}
    \begin{axis}[
    boxplots style,
    title={GTSRB based on poisoning rate},
    xlabel={Poisoning Rate},
    xmin=0.5, xmax=3.5,
    xtick={1,2,3},
    xticklabels={0.1,0.5,0.9},
    ylabel={Verfication Rate (0.001 Failure)},
    ymin=-0.00332509731804029, ymax=0.0737830620080433,
    ytick style={color=white!15!black}
    ]
    \addplot [boxplots box style,]
    table {%
    0.85 0.000557893346595061
    1.15 0.000557893346595061
    1.15 0.00395886597760594
    0.85 0.00395886597760594
    0.85 0.000557893346595061
    };
    \addplot [boxplots connectors style,]
    table {%
    1 0.000557893346595061
    1 0.000179819014963515
    };
    \addplot [boxplots connectors style,]
    table {%
    1 0.00395886597760594
    1 0.0076652750016832
    };
    \addplot [boxplots idonknow style,]
    table {%
    0.925 0.000179819014963515
    1.075 0.000179819014963515
    };
    \addplot [boxplots idonknow style,]
    table {%
    0.925 0.0076652750016832
    1.075 0.0076652750016832
    };
    \addplot [boxplots outpoints style,]
    table {%
    1 0.0102400947079447
    1 0.0116792318570615
    1 0.0100158977692191
    1 0.0117409482190389
    1 0.0128516088624823
    1 0.0143853641377161
    1 0.0290607027003359
    1 0.0223437556526394
    };
    \addplot [boxplots box style,]
    table {%
    1.85 0.00301161282070487
    2.15 0.00301161282070487
    2.15 0.0110048917422566
    1.85 0.0110048917422566
    1.85 0.00301161282070487
    };
    \addplot [boxplots connectors style,]
    table {%
    2 0.00301161282070487
    2 0.00074495506998012
    };
    \addplot [boxplots connectors style,]
    table {%
    2 0.0110048917422566
    2 0.0220543510827176
    };
    \addplot [boxplots idonknow style,]
    table {%
    1.925 0.00074495506998012
    2.075 0.00074495506998012
    };
    \addplot [boxplots idonknow style,]
    table {%
    1.925 0.0220543510827176
    2.075 0.0220543510827176
    };
    \addplot [boxplots outpoints style,]
    table {%
    2 0.0250936069359874
    2 0.0352674444237295
    2 0.0346782872574729
    2 0.0338713999390099
    2 0.0414036861073098
    2 0.0458424383917771
    2 0.0402393167582724
    2 0.0481491014475849
    2 0.0702781456750395
    };
    \addplot [boxplots box style,]
    table {%
    2.85 nan
    3.15 nan
    3.15 nan
    2.85 nan
    2.85 nan
    };
    \addplot [boxplots connectors style,]
    table {%
    3 nan
    3 nan
    };
    \addplot [boxplots connectors style,]
    table {%
    3 nan
    3 nan
    };
    \addplot [boxplots idonknow style,]
    table {%
    2.925 nan
    3.075 nan
    };
    \addplot [boxplots idonknow style,]
    table {%
    2.925 nan
    3.075 nan
    };
    \addplot [boxplots median style,]
    table {%
    0.85 0.00234715610604458
    1.15 0.00234715610604458
    };
    \addplot [boxplots meanpoints style,]
    table {%
    1 0.00431277640737046
    };
    \addplot [boxplots median style,]
    table {%
    1.85 0.00413342167021687
    2.15 0.00413342167021687
    };
    \addplot [boxplots meanpoints style,]
    table {%
    2 0.0124685873444853
    };
    \addplot [boxplots median style,]
    table {%
    2.85 nan
    3.15 nan
    };
    \addplot [boxplots meanpoints style,]
    table {%
    3 nan
    };
\end{axis}
\end{tikzpicture}

%% file: tex_plots/attack_vs_pois_cifar10.tex
\begin{tikzpicture}
    \begin{axis}[
    boxplots style,
    title={CIFAR10 based on poisoning rate},
    xlabel={Poisoning Rate},
    xmin=0.5, xmax=3.5,
    xtick={1,2,3},
    xticklabels={0.1,0.5,0.9},
    ylabel={Verfication Rate (0.001 Failure)},
    ymin=-0.00879455502637059, ymax=0.230077845690619,
    ]
    \addplot [boxplots box style,]
    table {%
    0.85 0.00498460852335905
    1.15 0.00498460852335905
    1.15 0.0231845561712766
    0.85 0.0231845561712766
    0.85 0.00498460852335905
    };
    \addplot [boxplots connectors style,]
    table {%
    1 0.00498460852335905
    1 0.0021710571965125
    };
    \addplot [boxplots connectors style,]
    table {%
    1 0.0231845561712766
    1 0.0349908949193132
    };
    \addplot [boxplots idonknow style,]
    table {%
    0.925 0.0021710571965125
    1.075 0.0021710571965125
    };
    \addplot [boxplots idonknow style,]
    table {%
    0.925 0.0349908949193132
    1.075 0.0349908949193132
    };
    \addplot [boxplots outpoints style,]
    table {%
    1 0.162461606349661
    1 0.11410395649095
    1 0.102263296686534
    };
    \addplot [boxplots box style,]
    table {%
    1.85 0.00743441119805707
    2.15 0.00743441119805707
    2.15 0.0680383556875147
    1.85 0.0680383556875147
    1.85 0.00743441119805707
    };
    \addplot [boxplots connectors style,]
    table {%
    2 0.00743441119805707
    2 0.00245114604631773
    };
    \addplot [boxplots connectors style,]
    table {%
    2 0.0680383556875147
    2 0.105388094624962
    };
    \addplot [boxplots idonknow style,]
    table {%
    1.925 0.00245114604631773
    2.075 0.00245114604631773
    };
    \addplot [boxplots idonknow style,]
    table {%
    1.925 0.105388094624962
    2.075 0.105388094624962
    };
    \addplot [boxplots outpoints style,]
    table {%
    2 0.219220009294392
    2 0.200150980820134
    2 0.189179758886118
    };
    \addplot [boxplots box style,]
    table {%
    2.85 0.00679932052027707
    3.15 0.00679932052027707
    3.15 0.021213495057815
    2.85 0.021213495057815
    2.85 0.00679932052027707
    };
    \addplot [boxplots connectors style,]
    table {%
    3 0.00679932052027707
    3 0.00206328136985622
    };
    \addplot [boxplots connectors style,]
    table {%
    3 0.021213495057815
    3 0.0367675113052911
    };
    \addplot [boxplots idonknow style,]
    table {%
    2.925 0.00206328136985622
    3.075 0.00206328136985622
    };
    \addplot [boxplots idonknow style,]
    table {%
    2.925 0.0367675113052911
    3.075 0.0367675113052911
    };
    \addplot [boxplots outpoints style,]
    table {%
    3 0.103802200938367
    3 0.0641843595514223
    };
    \addplot [boxplots median style,]
    table {%
    0.85 0.0102284158508223
    1.15 0.0102284158508223
    };
    \addplot [boxplots meanpoints style,]
    table {%
    1 0.0204085359134939
    };
    \addplot [boxplots median style,]
    table {%
    1.85 0.0419043262411808
    2.15 0.0419043262411808
    };
    \addplot [boxplots meanpoints style,]
    table {%
    2 0.0521870921058723
    };
    \addplot [boxplots median style,]
    table {%
    2.85 0.0179855891782808
    3.15 0.0179855891782808
    };
    \addplot [boxplots meanpoints style,]
    table {%
    3 0.0225269998846671
    };
\end{axis}
\end{tikzpicture}

%% file: 008relate.tex
TEEs~\cite{paper:intelsgxexplained,paper:arm_trustzone} are becoming more popular to provide privacy and integrity guarantee for applications across domains.~\cite{paper:sgx_big_matrix,paper:enclavedb,paper:sgx_zookeeper,paper:sgx_iron,paper:sgx_cloud_microservice,paper:sgx_ekiden,paper:wang2019towardsMemorySafeEnclave,paper:wang2019LanguageInterpreters,paper:baumanwang18sgxelide,paper:islam2020secure}. For deep learning inference, Slalom~\cite{Paper:Slalom} introduced a notion of mixed computation between SGX and GPU. It operates in two modes, one is pure-integrity and the other is privacy-preserving mode for inputs to the model (no privacy for the model parameters). It showed the great potential for outsourcing computation to a special hardware and verifying the computation inside enclave using Freivalds' scheme~\cite{paper:freivalds}. In contrast, \systemname{} aims at solving the integrity problem for the \textbf{training} phase. Authors in ~\cite{paper:sgx_origami} proposed \textit{Origami}; a privacy-preserving inference framework which partially computes some number of early layers in both enclave and GPU (similar to Slalom), while outsourcing the rest of the layers (input reconstruction as they claimed becomes infeasible) in plaintext solely onto the GPU.

For deep learning training using TEEs, Chiron~\cite{Paper:Chiron} was the first framework that proposed a distributed privacy-preserving solution using SGX. Yet in reality, private training within enclave and no other special hardware can take months for large architectures with potentially millions of records in the dataset. Especially, due to architectural limitations of SGX, special care must be taken to avoid allocating large enclaves (stay within PRM), and use partitioning scheme to enhance performance. TensorSCONE~\cite{paper:Tensorscone} is a porting of TensorFlow for SGX that supports both training and inference inside the enclave. It can support tensorflow graph definitions for models and it is integrable with current DNN ecosystem for applications that privacy is crucial. Myelin~\cite{Paper:Myelin} is a training framework on multi-source data where the parties do not trust each other. Myelin uses SGX to train the data in plaintext within enclave while applying differential privacy~\cite{paper:diff_privacy} measures to encounter potential privacy loss of the private inputs to other parties.

%% file: 099concl.tex
This paper introduced the \systemname{} system, which provides integrity in outsourced DNN training using TEEs. As our experimental investigation illustrates, \systemname{} scales up to realistic workloads by randomizing both mini-batch verification and matrix multiplication to achieve integrity guarantees with a high probability. We have further shown that random verification  in combination with hyperparameter adjustment (e.g., setting low clipping rates), can achieve 2\textbf{X}-20\textbf{X} performance improvements in comparison to pure TEE-based solutions while catching potential integrity violations with a very a high probability.

%% file: 0091adx.tex
\section{Model Signing By TEE}
\label{appendix:signature}
We assume, an honest and authenticated user will send her data encryption key $K_{client}$ (after remote-attestation) to the TEE. Next, the TEE decrypts and verifies the initial encrypted dataset using the $K_{client}$ and supplies the trainer (GPU) the plain-text of the training set. If the TEE fails to detect any violations of the protocol during training, it will sign the following message that certifies the final model where  $\mathcal{W}$ is the model parameters,  $ds$ is training dataset, $v\_num$  is the model version number, $SHA256$ is Sha 256 bit cryptographic hash function, 
\newline
\[SHA256(SHA256(\mathcal{W}) || v\_num || SHA256(ds)||Sig_{client}^{v})\]
\noindent with signature key $Sig_{SGX}^{s}$ of the enclave

\begin{table}[htb]
	\centering
	\caption{Symbols and Acronyms Description}
	\scalebox{0.8}{\begin{tabular}{l|c|c}
			Category & Symbol & Description \\ \hline
			\multirow{7}{*}{TEE}
			& $K_{SGX}^{session}$ & TEE's session key for learning task \\
			& $Sig_{SGX}^{s}$ & SGX signature signing key \\
			& $K_{client}$ & client's encryption key \\
			& $Sig_{client}^{v}$ & clients public key \\
			& PRM & Processor Reserved Memory \\
			& EPC & Enclave Page Cache \\
			\hline
			\multirow{3}{*}{Neural Network}
			& RMM & Randomized Matrix Multiplication \\
			& FV & Full Verification (No RMM)\\
			& DNN & Deep Neural Network \\ \hline
			
			\multirow{2}{*}{General}
			&$\mathcal{W}$& model parameters \\
			& $ds$ & training dataset \\ 
			& $v\_num$ & model version number \\ 
			\hline
	\end{tabular}}
	\label{tab:symbols}
\end{table}

\section{Deep Learning Training}
\label{subsection:training}
In the recent decade, Deep Neural Networks (DNN) gained an enormous attraction in solving problems related to computer vision and natural language processing~\cite{paper:alexnet,paper:VGG,paper:resnet,paper:GoogleNet,paper:inceptionv4}. In practice, these networks are stacks of layers that each perform a transformation $\mathcal{F}_{\mathcal{W}}^{l}(\cdot) \; \forall l \in |L|$ where $\mathcal{X}^{l+1} = \mathcal{F}_{\mathcal{W}}^{l}(\mathcal{X}^{l}) $ and $|L|$ is the number of layers.
The training task is to learn the correct parameters (point-estimates) $\mathcal{W^{*}}$ that optimizes (commonly minimizes) a task-specific (e.g. classification) loss function $\mathcal{L}$. 

The most common way of training the DNN's parameters is through mini-batch Stochastic Gradient Descent (SGD)~\cite{paper:optimizer_minibatch_sgd}. A randomly selected mini-batch of dataset is fed to the DNN and the value of objective function $\mathcal{L}$ is calculated. This is usually called the \textit{forward} pass. Next, to derive the partial gradients of $\mathcal{L}$ w.r.t $\mathcal{W}$ ($\mathcal{\nabla_{W}^{L}}$), a \textit{backward} pass is performed~\cite{paper:backprop1}. Finally, parameters are then updated according to the equation~\ref{eq:DNN_Loss} where $0 < \alpha < 1$ is called the learning rate. Depending on the complexity of the dataset and the task, it might require hundreds of passes (called \textit{epoch}) over the input dataset  for convergence.
\begin{eqnarray}
\mathcal{W}^{t+1} = \mathcal{W}^{t} - \alpha \mathcal{\nabla}_{\mathcal{W}^{t}}^{\mathcal{L}^{t}}
\label{eq:DNN_Loss}
\end{eqnarray}

\section{Gradient Clipping}
\label{subsection:gradclipping}
Gradient Clipping (GC) is a method that is mainly known to help mitigate the problem of exploding gradients during training~\cite{paper:gcexplodinggrad}. GC forces the gradients to fall within a narrow interval. There have been some efforts to analyze GC with respect to convergence. Authors in ~\cite{paper:gradclip_convergence} prove (assuming a fixed step size) that training with GC can be arbitrarily faster than training without it. Moreover, the theoretical analysis suggested that small clipping values can damage the training performance. However, in practice that is rarely the case. ~\cite{paper:gradclip_convergence_small} has an interesting theoretical analysis coupled with empirical evidence (symmetry of gradients distribution w.r.t SGD trajectory) that answers the gap between previous theoretical and practical observations.

\section{Integrity Proofs}

\subsection{Random Mini-Batch Verification}
\begin{definition}
Define random variable $X = \sum_{b=1}^{\ceil{ B\times p_{v}}} V_{b}$ to be the total number of random verification done. Here $V_b$ is $1$ if the batch is chosen for verification but failed the verification (due to malicious computation). Please note that we need to catch at least one deviation with probability greater than $p_{i}$, and invalidate the overall model learning.
\end{definition}
\begin{proof}
\label{proof:rand_mb_verf}
Proof of theorem ~\ref{theorem:rand_mb_verf}
\begin{eqnarray}
P(X \geq 1) = 1 - P(X = 0) &\geq& p_{i} \nonumber\\
1-p_{i} &\geq& P(X=0) \nonumber \\
1 - p_{i} &\geq& \binom{\ceil{ B\times p_{v}}}{0}p_{c}^0(1-p_{c})^{\ceil{ B\times p_{v}}} \nonumber\\
1 - p_{i} &\geq& (1-p_{c})^{\ceil{ B\times p_{v}}} \nonumber \\
\log(1 - p_{i}) &\geq& \ceil{ B\times p_{v}}\log(1-p_{c}) \nonumber \\
p_{v} &>& B^{-1}(\frac{\log(1 - p_{i})}{\log(1-p_{c})} - 1)
\label{eq:rand-sel}
\end{eqnarray}
\end{proof}
As shown in Figure. \ref{fig:random-verification-probs} it is only required to verify a much smaller subset of the batch computations inside the TEE to ensure a high probability of correct computation. For example, for large datasets such as Imagenet~\cite{Dataset:imagenet_cvpr09}, it is needed to verify less than \%1 of computation to have \emph{0.9999} guarantee (when corruption probability is \%0.5) on the correctness of computation outsourced to the GPU.

\subsection{Random Mini-Batch Verification with Randomized Matrix Multiplication}
\begin{definition}
Define random variable $V'_{b} = 1 \; if R_{b} = 1 \; \land \; MM\_verify(b) = 0 \; otherwise \; 0$ and $X = \sum_{b=1}^{\ceil{B\times p_{v}}} V'_{b}$. We need to detect at least one deviation with probability greater than $p_{i}$ while conducting random matrix multiplication verification.
\end{definition}
\begin{proof}
\label{proof:rand_mb_mm_verf}
Proof of theorem ~\ref{theorem:rand_mb_mm_verf}
\begin{eqnarray}
P(X \geq 1) &\geq& p_{i} \nonumber \\
1 - P(X = 0) &\geq& p_{i} \nonumber \\
1 - p_{i} &\geq& \binom{\ceil{ B\times p_{v}}}{0}(p_{c}(1-\alpha))^0((1-p_{c})+p_{c}\alpha)^{\ceil{ B\times p_{v}}} \nonumber \\
1 - p_{i} &\geq& ((1-p_{c})+p_{c}\alpha)^{\ceil{ B\times p_{v}}} \nonumber \\
\log(1 - p_{i}) &\geq& \ceil{ B\times p_{v}}\log((1-p_{c})+p_{c}\alpha) \nonumber \\
p_{v} &>& B^{-1}(\frac{\log(1 - p_{i})}{ \log((1+(\alpha-1)p_{c})}-1)
\label{eq:rand-sel-randmm}
\end{eqnarray}
\end{proof}

\section{Verification Probability Growth with Respect to Detection Probability}
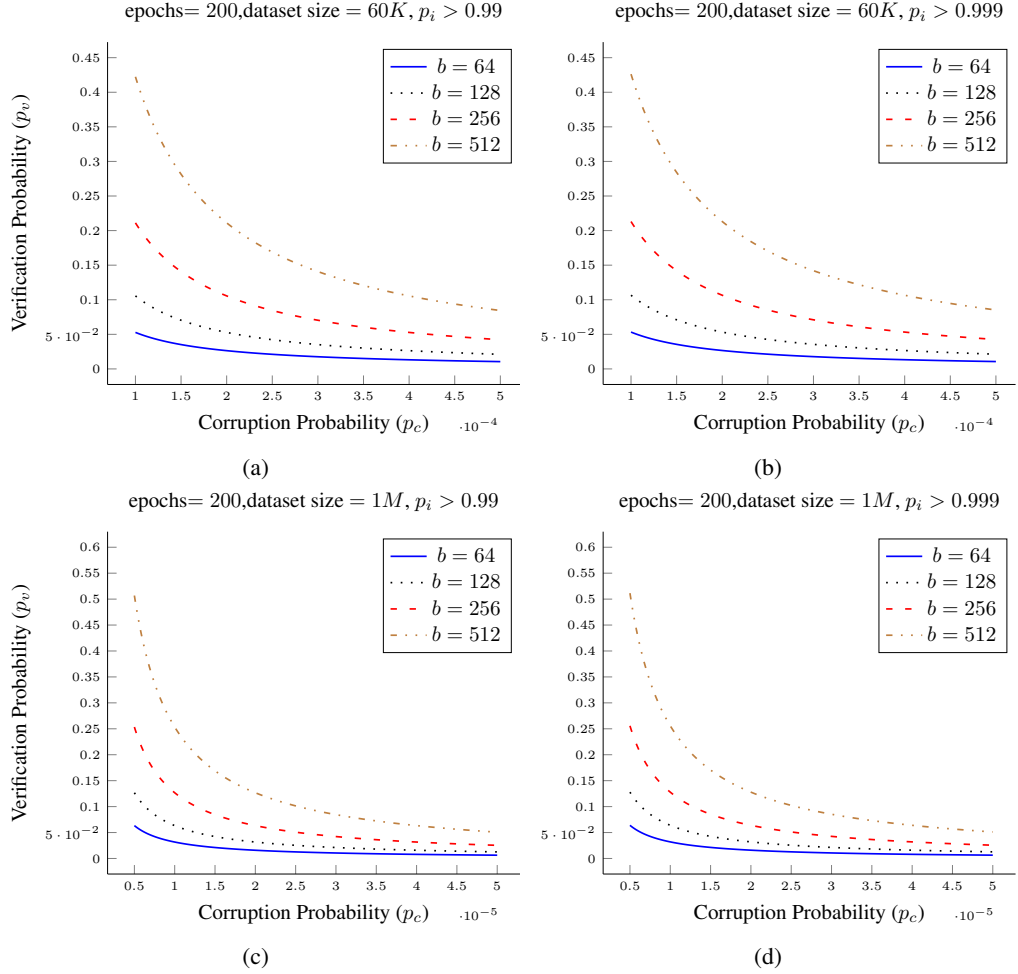
\begin{figure}[tb]
	\centering
	\subfloat[]{%
		\label{subfig:detect_0.99_ds_60K}
		\begin{tikzpicture}[scale=0.8]
		\begin{axis} [
		title={epochs$=200$,dataset size $=60K$, $p_{i}> 0.99$},
		ylabel=Verification Probability ($p_{v}$),
		xlabel=Corruption Probability ($p_{c}$),
		ymax=0.45,
		ymin=0.00001,
		xmin=0.00009,
		xmax=0.000501,
		xtick pos=left,
		ytick pos=left,
		enlargelimits=0.05,
		domain=0.0001:0.0005,
		scaled x ticks=true,
		scaled y ticks=false,
		try min ticks=10,
		xticklabel style={font=\tiny,
		},
		yticklabel style={font=\tiny},
        axis x line*=bottom,
        axis y line*=left,
		]
		\addplot [blue,samples=900,thick,solid]{
			(1/(200*(60000/64))) * ((ln{1.0-0.99})/(ln{1.0-x})-1.0)
		};
		\addlegendentry{{$b=64$}}
		\addplot [black,samples=900,thick,loosely dotted]{
			(1/(200*(60000/128))) * ((ln{1.0-0.99})/(ln{1.0-x})-1.0)
		};
		\addlegendentry{{$b=128$}}
		\addplot [red,samples=900,thick,loosely dashed]{
			(1/(200*(60000/256))) * ((ln{1.0-0.99})/(ln{1.0-x})-1.0)
		};
		\addlegendentry{{$b=256$}}
		\addplot [brown,samples=900,thick,loosely dashdotdotted]{
			(1/(200*(60000/512))) * ((ln{1.0-0.99})/(ln{1.0-x})-1.0)
		};
		\addlegendentry{{$b=512$}}
		\end{axis}
		\end{tikzpicture}
	}
	\subfloat[]{%
		\label{subfig:detect_0.999_ds_60K}
		\begin{tikzpicture}[scale=0.8]
		\begin{axis} [
		title={epochs$=200$,dataset size $=60K$, $p_{i}> 0.999$},
		xlabel=Corruption Probability ($p_{c}$),
		ymax=0.45,
		ymin=0.00001,
		xmin=0.00009,
		xmax=0.000501,
		xtick pos=left,
		ytick pos=left,
		enlargelimits=0.05,
		domain=0.0001:0.0005,
		scaled x ticks=true,
		scaled y ticks=false,
		try min ticks=10,
		xticklabel style={font=\tiny,
		},
		yticklabel style={font=\tiny},
        axis x line*=bottom,
        axis y line*=left,
		]
		\addplot [blue,samples=900,thick,solid]{
			(1/(200*(60000/64))) * ((ln{1.0-0.999})/(ln{1.0-x})-1.0)
		};
		\addlegendentry{{$b=64$}}
		\addplot [black,samples=900,thick,loosely dotted]{
			(1/(200*(60000/128))) * ((ln{1.0-0.999})/(ln{1.0-x})-1.0)
		};
		\addlegendentry{{$b=128$}}
		\addplot [red,samples=900,thick,loosely dashed]{
			(1/(200*(60000/256))) * ((ln{1.0-0.999})/(ln{1.0-x})-1.0)
		};
		\addlegendentry{{$b=256$}}
		\addplot [brown,samples=900,thick,loosely dashdotdotted]{
			(1/(200*(60000/512))) * ((ln{1.0-0.999})/(ln{1.0-x})-1.0)
		};
		\addlegendentry{{$b=512$}}
		\end{axis}
		\end{tikzpicture}
	}
	
	\subfloat[]{%
		\label{subfig:detect_0.99_ds_1M}
		\begin{tikzpicture}[scale=0.8]
		\begin{axis} [
		title={epochs$=200$,dataset size $=1M$, $p_{i}> 0.99$},
		ylabel=Verification Probability ($p_{v}$),
		xlabel=Corruption Probability ($p_{c}$),
		ymax=0.6,
		ymin=0.00001,
		xmin=0.000004,
		xmax=0.0000505,
		xtick pos=left,
		ytick pos=left,
		scaled y ticks=false,
		scaled x ticks=true,
		enlargelimits=0.05,
		domain=0.000005:0.00005,
		try min ticks=10,
		xticklabel style={font=\tiny,
		},
		yticklabel style={font=\tiny},
		axis x line*=bottom,
        axis y line*=left,
		]
		\addplot [blue,samples=2000,thick,solid]{
			(1/(200*(1000000/64))) * ((ln{1.0-0.99})/(ln{1.0-x})-1.0)
		};
		\addlegendentry{{$b=64$}}
		\addplot [black,samples=2000,thick,loosely dotted]{
			(1/(200*(1000000/128))) * ((ln{1.0-0.99})/(ln{1.0-x})-1.0)
		};
		\addlegendentry{{$b=128$}}
		\addplot [red,samples=2000,thick,loosely dashed]{
			(1/(200*(1000000/256))) * ((ln{1.0-0.99})/(ln{1.0-x})-1.0)
		};
		\addlegendentry{{$b=256$}}
		\addplot [brown,samples=2000,thick,loosely dashdotdotted]{
			(1/(200*(1000000/512))) * ((ln{1.0-0.99})/(ln{1.0-x})-1.0)
		};
		\addlegendentry{{$b=512$}}
		\end{axis}
		\end{tikzpicture}
	}
	\subfloat[]{%
		\label{subfig:detect_0.999_ds_1M}
		\begin{tikzpicture}[scale=0.8]
		\begin{axis} [
		title={epochs$=200$,dataset size $=1M$, $p_{i}> 0.999$},
		xlabel=Corruption Probability ($p_{c}$),
		ymax=0.6,
		ymin=0.00001,
		xmin=0.000004,
		xmax=0.0000505,
		xtick pos=left,
		ytick pos=left,
		scaled y ticks=false,
		scaled x ticks=true,
		enlargelimits=0.05,
		domain=0.000005:0.00005,
		try min ticks=10,
		xticklabel style={font=\tiny,
		},
		yticklabel style={font=\tiny},
		axis x line*=bottom,
        axis y line*=left,
		]
		\addplot [blue,samples=2000,thick,solid]{
			(1/(200*(1000000/64))) * ((ln{1.0-0.999})/(ln{1.0-x})-1.0)
		};
		\addlegendentry{{$b=64$}}
		\addplot [black,samples=2000,thick,loosely dotted]{
			(1/(200*(1000000/128))) * ((ln{1.0-0.999})/(ln{1.0-x})-1.0)
		};
		\addlegendentry{{$b=128$}}
		\addplot [red,samples=2000,thick,loosely dashed]{
			(1/(200*(1000000/256))) * ((ln{1.0-0.999})/(ln{1.0-x})-1.0)
		};
		\addlegendentry{{$b=256$}}
		\addplot [brown,samples=2000,thick,loosely dashdotdotted]{
			(1/(200*(1000000/512))) * ((ln{1.0-0.999})/(ln{1.0-x})-1.0)
		};
		\addlegendentry{{$b=512$}}
		\end{axis}
		\end{tikzpicture}
	}
	\caption{Required verification probability with respect to batch corruption probability and the desired integrity probability for a fixed 200 epochs and different SGD batch size. 
	}
	\label{fig:random-verification-probs}
\end{figure}

Fig.~\ref{fig:random-verification-probs} shows how verification probability changes with respect to the probability that a batch step is maliciously manipulated by the attacker. First row shows the verification probability for a dataset with 60K samples. Second row depicts the required for much bigger dataset (1M samples) over different mini-batch sizes. The smaller the mini-batch size is, there is a higher chance for detecting malicious behavior.

\section{Experimental Evaluation}
\subsection{Enclave Heap Size Impact on TEE Performance}
\begin{figure}[t]
	\begin{center}
	\subfloat[Available heap with respect to throughput]{%
		\begin{tikzpicture}%
		\begin{axis}[sharp plot,
		xlabel={(SGX Max Heap Size, Blocking Size) MB},
		ylabel={Throughput (Images/Sec)},
		ymin=0.1,
		ymax=1.9,
		xticklabels={{(100,32)},{(150,48)},{(180,64)},{(200,80)},{(220,96)}},
		xtick=data,
		yticklabels={0.1,0.25,0.5,0.75,1.0,1.25,1.5,1.75,2.0},
		ytick={0.1,0.25,0.5,0.75,1.0,1.25,1.5,1.75,2.0},
		legend columns = 4,
		legend style = {
			at={(.55, 0.68)}, 
						anchor=north, 
						inner sep=1pt,
						style={column sep=0.05cm},
			nodes={
			scale=0.45,
			transform shape},
			cells={align=left,anchor=west},
		},
		axis lines*=left,
        width=.5\textwidth,
        height=.3\textwidth,
        xticklabel style={font=\tiny},
        yticklabel style={font=\tiny},
        xlabel style={font=\tiny},
        ylabel style={font=\tiny},
		]
		
		\addlegendimage{black,only marks,mark=square*} %
		\addlegendentry{VGG19}
		
		\addlegendimage{black,only marks,mark=diamond} %
		\addlegendentry{VGG16}
		
		\addlegendimage{black,only marks,mark=triangle*} %
		\addlegendentry{ResNet152}
		
		\addlegendimage{black,only marks,mark=pentagon} %
		\addlegendentry{ResNet34}
		
		\addlegendimage{red,line legend}
		\addlegendentry{\makebox[0pt][l]{SGX}}
		
		\addlegendimage{empty legend}
		\addlegendentry{}
		
		\addlegendimage{blue,line legend,loosely dashdotdotted}
		\addlegendentry{\makebox[0pt][l]{SGX RMM}}
		
		\addplot[color=red,mark=triangle*,mark options={solid}] coordinates {
			(1,0.4286207975)
			(2,0.42832118)
			(3,0.4317611809)
			(4,0.4054276782)
			(5,0.4099751594)
		};
		
		\addplot[color=blue,loosely dashdotdotted,mark=triangle*,mark options={solid}] coordinates {
			(1,0.5091803588)
			(2,0.5128553853)
			(3,0.5189976453)
			(4,0.4832300157)
			(5,0.4842871703)
		};
		
		\addplot[color=red,mark=pentagon,mark options={solid}] coordinates {
			(1,1.413340008)
			(2,1.426687564)
			(3,1.410301974)
			(4,1.428368046)
			(5,1.426795196)
		};
		
		\addplot[color=blue,loosely dashdotdotted,mark=pentagon,mark options={solid}] coordinates {
			(1,1.752690016)
			(2,1.760609633)
			(3,1.780741377)
			(4,1.776867528)
			(5,1.786623001)
		};
		
		\addplot[color=red,mark=square*,mark options={solid}] coordinates {
			(1,0.2726836983)
			(2,0.2901144753)
			(3,0.2909748658)
			(4,0.2663475845)
			(5,0.2112050663)
		};
		
		\addplot[color=blue,loosely dashdotdotted,mark=square*,mark options={solid}] coordinates {
			(1,0.6311147749)
			(2,0.5718062852)
			(3,0.5703824493)
			(4,0.5254294729)
			(5,0.4210763144)
		};
		
		\addplot[color=red,mark=diamond,mark options={solid}] coordinates {
			(1,0.3077821949)
			(2,0.3310894556)
			(3,0.3318761619)
			(4,0.3113047857)
			(5,0.2433932361)
		};
		
		\addplot[color=blue,loosely dashdotdotted,mark=diamond,mark options={solid}] coordinates {
			(1,0.7007663515)
			(2,0.6389719644)
			(3,0.6368137533)
			(4,0.5932088652)
			(5,0.4718251353)
		};
		
		\end{axis}
		\end{tikzpicture}
		\label{subfig:throughput_heap}
	}
	\subfloat[Available heap with respect to time spent on matrix-matrix(vector) multiplication]{%
 		\begin{tikzpicture}%
		\begin{axis}[sharp plot,
		xlabel={(SGX Max Heap Size, Blocking Size) MB},
		ylabel={Time (Sec)},
		xticklabels={{(100,32)},{(150,48)},{(180,64)},{(200,80)},{(220,96)}},
		xtick=data,
		axis lines*=left,
		width=.5\textwidth,
        height=.3\textwidth,
        xticklabel style={font=\tiny},
        yticklabel style={font=\tiny},
        xlabel style={font=\tiny},
        ylabel style={font=\tiny},
		]

		\addplot[color=red,mark=triangle*,mark options={solid}] coordinates {
			(1,38.185016)
			(2,38.072908)
			(3,37.711327)
			(4,40.612592)
			(5,40.142905)
		};
		
		\addplot[color=blue,loosely dashdotdotted,mark=triangle*,mark options={solid}] coordinates {
			(1,21.480757)
			(2,21.039984)
			(3,20.193895)
			(4,23.620996)
			(5,23.635319)
		};
		
		\addplot[color=red,mark=pentagon,mark options={solid}] coordinates {
			(1,20.412225)
			(2,20.198708)
			(3,20.33375)
			(4,20.18655)
			(5,20.235605)
		};
		
		\addplot[color=blue,loosely dashdotdotted,mark=pentagon,mark options={solid}] coordinates {
			(1,13.448346)
			(2,13.51716)
			(3,13.172704)
			(4,13.276503)
			(5,13.091452)
		};
		
		\addplot[color=red,mark=square*,mark options={solid}] coordinates {
			(1,165.139311)
			(2,148.394364)
			(3,148.479209)
			(4,154.785324)
			(5,203.783577)
		};
		
		\addplot[color=blue,loosely dashdotdotted,mark=square*,mark options={solid}] coordinates {
			(1,39.002411)
			(2,45.289763)
			(3,44.895975)
			(4,44.999765)
			(5,55.11462)
		};
		
		\addplot[color=red,mark=diamond,mark options={solid}] coordinates {
			(1,146.94794)
			(2,130.328297)
			(3,129.872369)
			(4,130.306258)
			(5,177.165922)
		};
		
		\addplot[color=blue,loosely dashdotdotted,mark=diamond,mark options={solid}] coordinates {
			(1,35.667228)
			(2,40.37162)
			(3,39.963483)
			(4,40.954017)
			(5,49.954216)
		};
		\end{axis}
		\end{tikzpicture}
		\label{subfig:gemm_heap}
	}
	\end{center}
	\caption{The impact of increasing TEE heap size on (\protect{\subref{subfig:throughput_heap}}) overall throughput and (\protect{\subref{subfig:gemm_heap}}) the time spent in matrix multiplication routine . VGG shows significant reduction in performance as opposed to ResNet.}
	\label{fig:heap_block_impact_performance}
\end{figure}
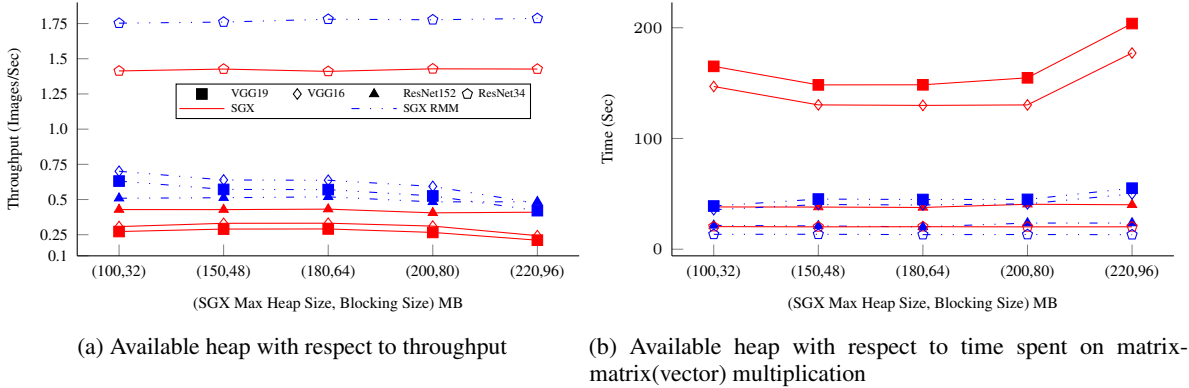
As shown in Figure~\ref{fig:heap_block_impact_performance}, which depicts the impact of the heap size on the performance of DNNs for a single SGD step, it can be seen that increasing the heap size way beyond the available hardware limit (around 92MB) can have a negative impact on performance especially in the case of the VGG architecture. This result is mainly due to 1) driver level paging, which needs to evict enclave pages that require an extra level of encryption/decryption and 2) extra bookkeeping for the evicted pages.
\subsection{TEE Performance on CIFAR10}
\label{subsec:appdx_tee_cifar10}
\begin{table}[t]
	\centering
	\caption{TEE Architectures Used}
	\begin{tabular}{|c|c|c|c|}
		\hline
		\textbf{Arch} & \textbf{FC1} & \textbf{FC2} & \textbf{FC3} \\
		\hline
		VGG11 & (128,10) & (128,64,10) & (256,128,10) \\
		\hline
		VGG13 & (128,10) & (128,64,10) & (256,128,10) \\
		\hline
		VGG16 & (128,10) & (128,64,10) & (256,128,10) \\
		\hline
	\end{tabular}
	\label{tab:sgx_cifar_archs}
\end{table}
\begin{figure}[t]
    \begin{center}
	    \begin{tikzpicture}%
		\begin{groupplot}[
		group style={
		    group name=teeperf,
		    group size=9 by 1,
		    /pgf/bar width=.2,
		    ylabels at=edge left,
		    xlabels at=edge bottom,
		    y descriptions at=edge left,
		  horizontal sep=0.5cm,
		    },
        ybar=\pgflinewidth,
		ylabel=Throughput (Images/Sec),
		ylabel style={font=\small},
		xticklabels={SGX,SGX\textsuperscript{RMM}},
		xticklabel style={font=\tiny},
		yticklabel style={font=\tiny},
		xtick = data,
		nodes near coords,
		nodes near coords style={font=\tiny,rotate=90,anchor=west},
		ymax=60,
		ymin=5,
		axis lines*=left,
		width=0.18\textwidth,
		height=0.40\textwidth,
		clip=false,
		ytick align=outside,
        enlarge x limits={abs=.22},
        legend entries = {Forward,Backward,Overall},
        legend columns=-1,
        legend to name=legendnamed2,
		]
		\nextgroupplot[
		xlabel=VGG11-FC1,xlabel style={font=\scriptsize}]
		\addplot coordinates {
			(1,59.84348124) (2,45.75329039)
		};
		\addplot coordinates {
			(1,17.02189002) (2,34.59035027)
		};
		\addplot coordinates {
			(1,13.252380619) (2,17.07460978)
		};
		
		\nextgroupplot[
		xlabel=VGG11-FC2,y axis line style={draw opacity=0},ytick style={draw=none},xlabel style={font=\scriptsize}]
		\addplot coordinates {
			(1,57.11175164) (2,45.30102709)
		};
		\addplot coordinates {
			(1,16.9941199) (2,31.43602977)
		};
		\addplot coordinates {
			(1,13.0969913) (2,15.73407254)
		};
		
		\nextgroupplot[
		xlabel=VGG11-FC3,y axis line style={draw opacity=0},ytick style={draw=none},,xlabel style={font=\scriptsize}]
		\addplot coordinates {
			(1,57.41802792) (2,53.27388855)
		};
		\addplot coordinates {
			(1,17.09887867) (2,32.76016691)
		};
		\addplot coordinates {
			(1,13.17531736) (2,17.30244361)
		};
		
		\nextgroupplot[
		xlabel=VGG13-FC1,y axis line style={draw opacity=0},ytick style={draw=none},xlabel style={font=\scriptsize}]
		\addplot coordinates {
		    (1,50.90926736) (2,41.35537058)
		};
		\addplot coordinates {
		    (1,16.26040634) (2,24.89223413)
		};
		\addplot coordinates {
		    (1,12.32409402) (2,12.75320848)
		};
		
		\nextgroupplot[
		xlabel=VGG13-FC2,y axis line style={draw opacity=0},ytick style={draw=none},xlabel style={font=\scriptsize}]
		\addplot coordinates {
			(1,47.98793703) (2,42.32061103)
		};
		\addplot coordinates {
			(1,16.10680017) (2,31.61707767)
		};
		\addplot coordinates {
			(1,12.05921338) (2,16.76332426)
		};
		
		\nextgroupplot[
		xlabel=VGG13-FC3,y axis line style={draw opacity=0},ytick style={draw=none},xlabel style={font=\scriptsize}]
		\addplot coordinates {
			(1,48.31360968) (2,40.75428541)
		};
		\addplot coordinates {
			(1,16.04266748) (2,32.65643517)
		};
		\addplot coordinates {
			(1,12.04356761) (2,16.9894733)
		};
		
		\nextgroupplot[
		xlabel=VGG16-FC1,y axis line style={draw opacity=0},ytick style={draw=none},xlabel style={font=\scriptsize}]
		\addplot coordinates {
		    (1,40.17692915) (2,34.02410236)
		};
		\addplot coordinates {
		    (1,10.74517991) (2,19.2535721)
		};
		\addplot coordinates {
		    (1,8.477817195) (2,9.824769857)
		};
		
		\nextgroupplot[
		xlabel=VGG16-FC2,y axis line style={draw opacity=0},ytick style={draw=none},xlabel style={font=\scriptsize}]
		\addplot coordinates {
			(1,39.29785781) (2,30.6699589)
		};
		\addplot coordinates {
			(1,10.78273744) (2,19.87296209)
		};
		\addplot coordinates {
			(1,8.461131117) (2,9.95890396)
		};
		
		\nextgroupplot[
		xlabel=VVG16-FC3,y axis line style={draw opacity=0},ytick style={draw=none},xlabel style={font=\scriptsize}]
		\addplot coordinates {
			(1,38.10416309) (2,34.39118468)
		};
		\addplot coordinates {
			(1,10.71732568) (2,17.43162683)
		};
		\addplot coordinates {
			(1,8.364651219) (2,9.011844239)
		};
        \end{groupplot}
        \node [above] at (current bounding box.north) {Throughput Performance (CIFAR10)};
		\end{tikzpicture}
		\ref{legendnamed2}
	\caption{Throughput of SGD training step for VGG19,VGG16, ResNet152, and Resnet34 on \mbox{ImageNet} dataset. Randomized Matrix Multiplication can make verification twice faster in case of VGG architecture.}
	\label{fig:vgg_cifar10_multiconv_multifc_performance}
	\end{center}
\end{figure}
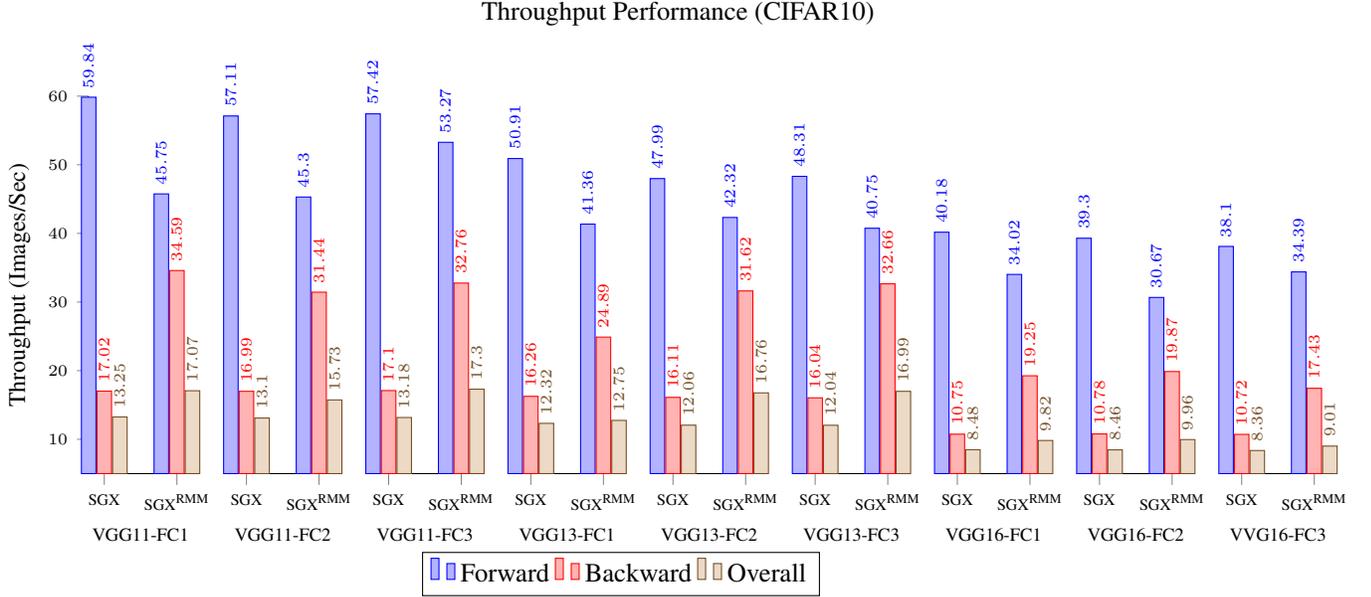
Figure~\ref{fig:vgg_cifar10_multiconv_multifc_performance} shows throughput performance for CIFAR10 dataset and 9 different VGG architectures (Table~\ref{tab:sgx_cifar_archs}). We chose three VGG(11,13,16) architectures adapted for CIFAR10 image inputs with custom fully connected layers attached to its end. For CIFAR10, we generally do not benefit from randomized matrix multiplication scheme as well as ImageNet. Mainly it is because most of the operations and network layers fit well within the hardware memory limit. Therefore, since dimensions of MM operations are not too big, it does not improve significantly to use randomized matrix multiplications.

\subsection{Impact of Gradient Clipping for Honest Trainers}
\begin{figure}[t]
    \begin{center}
    \subfloat[]{%
    \label{fig:no_attack_clean_refruns_meanstd}
    \input{tex_plots/mean_std_ref_acc}
    }
    
    \subfloat[]{%
    \label{fig:no_attack_clean_runs_best}
    \input{tex_plots/best_gaps}
    }
    
    \subfloat[]{%
    \label{fig:no_attack_clean_runs_worst}
    \input{tex_plots/worst_gaps}
    }
    \end{center}
    \caption{~\ref{fig:no_attack_clean_refruns_meanstd} Reference Models (no gradient clipping) mean/std on test accuracy of 5 repeats for two different learning rates. Each configuration had 5 repeats and a reference model (no attack and unbounded updates).~\ref{fig:no_attack_clean_runs_best} For each run configuration the test accuracy difference $diff_{lr,clip}$ is defined as $max\left( acc_{lr,clip}^{rep}-acc_{ref}^{rep}\right) \quad \forall rep \in [1,5]$.~\ref{fig:no_attack_clean_runs_worst} $min\left( acc_{lr,clip}^{rep}-acc_{ref}^{rep}\right) \quad \forall rep \in [1,5]$}
    \label{fig:no_attack_clean_runs}
\end{figure}
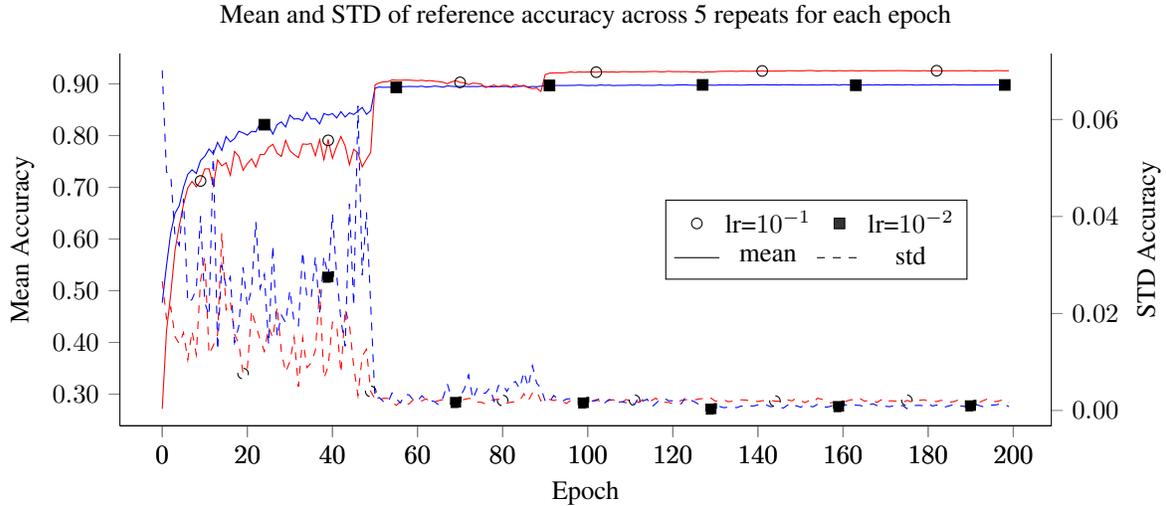
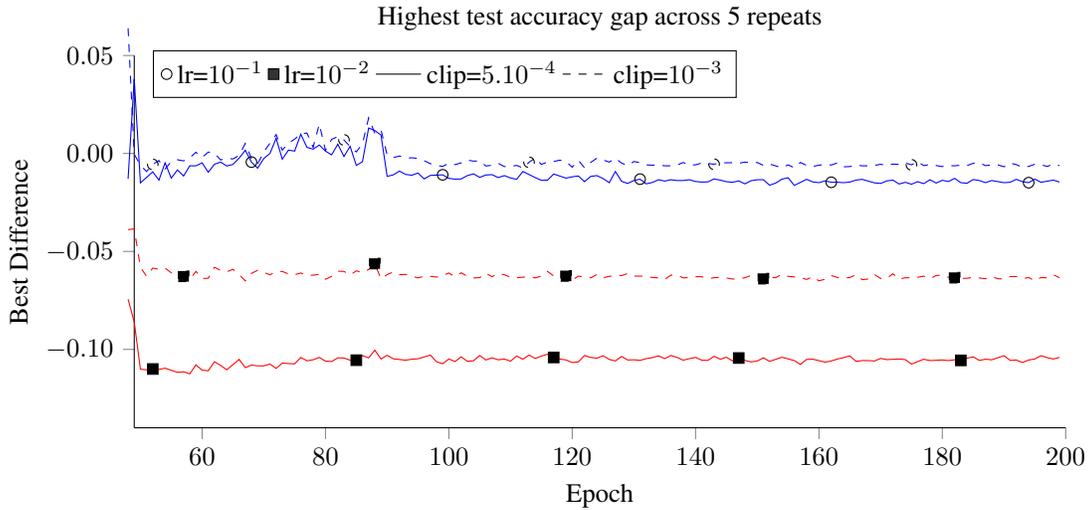
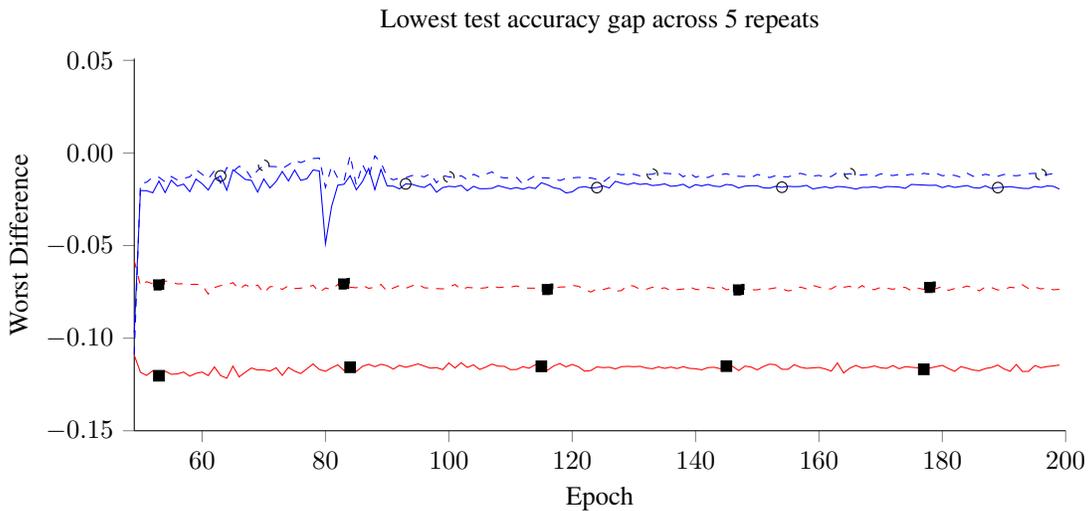

One important question is whether the gradient clipping used to prevent attacker to change parameters in a given mini-batch update would have performance impact during training session where there is no attack.
Six experiment configurations were repeated $5$ times each with different randomness (initialization, batch order, etc.). Initial learning rates are set $ \in (0.1,0.01)$ and clipping thresholds are set $\in (nil,0.001,0.0005)$.
In total, there are 30 ResNet56 (complex architecture with state-of-the-art performance) models trained on the CIFAR10 dataset (with no attack) for $200$ epochs. It is shown that the clipping rate have very little impact on the test set accuracy, given an appropriate initial learning rate is chosen.
Usually, for \textit{SGD}~\cite{paper:optimizer_minibatch_sgd} optimizer with momentum, a value of $0.1$ is chosen, (and for \textit{Adam} optimizer a value less than $0.001$).
For these experiments, we used the configuration with unbounded gradient updates as the main reference point. For learning decay schedule, we used a fixed decay by tenfold at epochs (50, 90, 130,160).

In figure~\ref{fig:no_attack_clean_refruns_meanstd} describes the mean and standard deviation (dashed lines) of test accuracy taken for 5 repetitions at each epoch for the two learning rate configurations. As it can be seen both models start to take giant leaps toward convergence at the first two learning decays enforced by the scheduler. Please note that these are reference runs that no gradient clipping is applied during the update step. Toward the end of training, the setting with the higher initial learning rate slightly performs better in terms of accuracy (the y-axis is not in \% scale).

In figure~\ref{fig:no_attack_clean_runs_best}, for each composition of learning rate, and clipping value, the highest difference (accuracy rate) with respect to reference run is plotted. The plot shows test accuracy is not influenced so much by the clipping value, rather, it is highly dependent on the learning rate value. when $lr=0.1$, both clipping values can achieve values that are close to the reference runs that has no gradient clipping, however, slightly smaller (most epochs it is negative, except jumps in the start).
In figure~\ref{fig:no_attack_clean_runs_worst}, the opposite of the previous measure is plotted. Again, by the end of the training, the gaps are significantly tightened for the case where a better learning is chosen. Therefore, having a smaller clipping value is not really impacting the performance in any considerable way.

Overall, figure~\ref{fig:no_attack_clean_runs}, shows that clipping does not really impact the learning task negatively, once a good learning rate is chosen. One can observe that if the trainer choose an acceptable learning rate for the task, small clipping values such as $0.001$ or $0.0005)$ does not impede the learning task. Once the model passes the first learning rate decay schedule, all the configuration behave the same in terms of their test performance compared to their reference model (no gradient clipping limit).
\subsection{All Backdoor Trigger Examples}
\begin{figure}[!htb]
	\begin{center}
	\subfloat[\tiny Large with multiple color variations]{%
	    \label{fig:appdx_CIFAR10_Trigger_Example:trigger1}
		\scalebox{.95}{\includegraphics{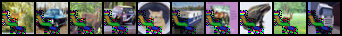}}
	}%
	
	\subfloat[\tiny Small with low color variations in two separate corners]{%
	\label{fig:appdx_CIFAR10_Trigger_Example:trigger2}
	\scalebox{.95}{\includegraphics{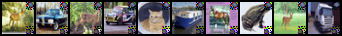}}
	}%
	
    \subfloat[\tiny Gray scale]{%
	\label{fig:appdx_CIFAR10_Trigger_Example:trigger3}
	\scalebox{.95}{\includegraphics{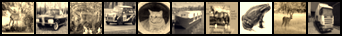}}
	}%
	
    \subfloat[\tiny Instagram Kelvin Filter]{%
	\label{fig:appdx_CIFAR10_Trigger_Example:trigger4}
	\scalebox{.95}{\includegraphics{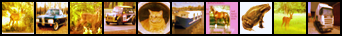}}
	}%

    \subfloat[\tiny Color Rotation Filter]{%
	\label{fig:appdx_CIFAR10_Trigger_Example:trigger5}
	\scalebox{.95}{\includegraphics{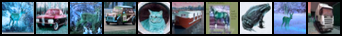}}
	}%
	
    \subfloat[\tiny Mix of Rotation and Instagram Nashville Filters]{%
	\label{fig:appdx_CIFAR10_Trigger_Example:trigger6}
	\scalebox{.95}{\includegraphics{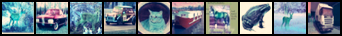}}
	}%
	\end{center}
	\caption[]{All examples of triggers on CIFAR10 images}
	\label{fig:appdx_CIFAR10_Trigger_Example}
\end{figure}
Figure~\ref{fig:appdx_CIFAR10_Trigger_Example} shows six different backdoor trigger patterns that are used to conduct the attacks.

\section{Matrix Multiplication Ops of Common DNN Layers}
\label{section:MM_OPS_APX}
\begin{table}[htb]
\centering
\caption{Matrix Multiplication Operations}
\label{tab:MM-OPS-Verification}
\begin{adjustbox}{width=1\textwidth}
\begin{tabular}{|c|c|l|c|
}
\hline
\textbf{Layer Type} & \textbf{Pass} & \textbf{Computation} & \textbf{Verification} 
\\
\hline
\multirow{3}{*}{Fully Connected}&Forward& $\mathcal{O}_{[B][O]} = \mathcal{I}_{[B][I]} \times (\mathcal{W}_{[O][I]})^\intercal $ &
$\begin{aligned}
\Upsilon_{[I][1]} & = (\mathcal{W}_{[O][I]})^\intercal \times \mathcal{R}_{[O][1]} \\
\mathcal{Z}_{[B][1]} & = \mathcal{I}_{[B][I]} \times \Upsilon_{[I][1]} \\
\mathcal{Z^{\prime}}_{[B][1]} & = \mathcal{O}_{[B][O]} \times \mathcal{R}_{[O][1]}
\end{aligned}$
\\
\cline{2-4}
&\makecell{Backward Parameters \\ Gradient}&
$\nabla_{[O][I]}^{\mathcal{W}} = (\nabla_{[B][O]}^{\mathcal{O}})^\intercal \times \mathcal{I}_{[B][I]}$ &
$\begin{aligned}
\Upsilon_{[B][1]} & = \mathcal{I}_{[B][I]} \times \mathcal{R}_{[I][1]} \\
\mathcal{Z}_{[O][1]} & = (\nabla_{[B][O]}^{\mathcal{O}})^\intercal \times \Upsilon_{[B][1]} \\
\mathcal{Z^{\prime}}_{[O][1]} & = \nabla_{[O][I]}^{\mathcal{W}} \times \mathcal{R}_{[I][1]}
\end{aligned}$
\\
\cline{2-4}
&\makecell{Backward Inputs \\ Gradient}&
$\nabla_{[B][I]}^{\mathcal{I}} = \nabla_{[B][O]}^{\mathcal{O}} \times \mathcal{W}_{[O][I]}$ &
$\begin{aligned} \Upsilon_{[O][1]} & = \mathcal{W}_{[O][I]} \times \mathcal{R}_{[I][1]} \\
\mathcal{Z}_{[B][1]} & = \nabla_{[B][O]}^{\mathcal{O}} \times \Upsilon_{[O][1]} \\
\mathcal{Z^{\prime}}_{[B][1]} & = \nabla_{[B][I]}^{\mathcal{I}} \times \mathcal{R}_{[I][1]}
\end{aligned}$
\\
\hline
\multirow{3}{*}{Convolutional}&Forward& $\mathcal{O}_{[f] [w_{o} . h_{o}]} = \mathcal{W}_{[f] [k^{2} . C_{i}]} \times \mathcal{I}_{[k^{2} . C_{i}][w_{o} . h_{o}]}$ &
$\begin{aligned}
\Upsilon_{[1][k^{2} . C_{i}]} & = \mathcal{R}_{[1][f]} \times \mathcal{W}_{[f] [k^{2} . C_{i}]} \\
\mathcal{Z}_{[1][w_{o} . h_{o}]} & = \Upsilon_{[1][k^{2} . C_{i}]} \times \mathcal{I}_{[k^{2} . C_{i}][w_{o} . h_{o}]} \\
\mathcal{Z^{\prime}}_{[1][w_{o} . h_{o}]} & = \mathcal{R}_{[1][f]} \times \mathcal{O}_{[f] [w_{o} . h_{o}]}
\end{aligned}$
\\
\cline{2-4}
&\makecell{Backward Parameters \\ Gradient}&
$\nabla_{[f][k^{2} . C_{i}]}^{\mathcal{W}} = \nabla_{[f][w_{o} . h_{o}]}^{\mathcal{O}} \times (\mathcal{I}_{[k^{2} . C_{i}][w_{o} . h_{o}]})^\intercal $ &
$\begin{aligned}
\Upsilon_{[w_{o} . h_{o}][1]} & = (\mathcal{I}_{[k^{2} . C_{i}][w_{o} . h_{o}]})^\intercal \times \mathcal{R}_{[k^{2} . C_{i}][1]} \\
\mathcal{Z}_{[f][1]} & = \nabla_{[f][w_{o} . h_{o}]}^{\mathcal{O}} \times \Upsilon_{[w_{o} . h_{o}][1]} \\
\mathcal{Z^{\prime}}_{[f][1]} & = \nabla_{[f][k^{2} . C_{i}]}^{\mathcal{W}} \times \mathcal{R}_{[k^{2} . C_{i}][1]}
\end{aligned}$
\\
\cline{2-4}
&\makecell{Backward Inputs \\ Gradient}&$\nabla_{[k^{2} . C_{i}][w_{o} . h_{o}]}^{\mathcal{I}} = (\mathcal{W}_{[f] [k^{2} . C_{i}]})^\intercal \times \nabla_{[f][w_{o} . h_{o}]}^{\mathcal{O}}$ &
$\begin{aligned}
\Upsilon_{[1][f]} & = \mathcal{R}_{[1][k^{2} . C_{i}]} \times (\mathcal{W}_{[f] [k^{2} . C_{i}]})^\intercal \\
\mathcal{Z}_{[1][w_{o} . h_{o}]} & = \Upsilon_{[1][f]} \times \nabla_{[f][w_{o} . h_{o}]}^{\mathcal{O}} \\
\mathcal{Z^{\prime}}_{[1][w_{o} . h_{o}]} & = \mathcal{R}_{[1][k^{2} . C_{i}]} \times \nabla_{[k^{2} . C_{i}][w_{o} . h_{o}]}^{\mathcal{I}}
\end{aligned}$
\\
\hline
\end{tabular}
\end{adjustbox}
\end{table}
Table.~\ref{tab:MM-OPS-Verification} shows common MM operations in DNNs. Connected and convolutional layers use MM routines to compute feed forward output, parameter gradients, and input gradients.

\section{\systemname{} Blocking of Big Matrices}

By default, \systemname{} allocates/releases resources on a per layer basis. In the event that even for a single sample, it is not possible to satisfy the memory requirements of network (either large network or large inputs), \systemname{} breaks each layer even further.

For convolutional layers, the main memory bottleneck is \textit{im2col}\footnote{extracts redundant patches from the input image and lays in columnar format} which converts the layer's input (for each sample) of size $c_i \cdot w_i \cdot h_i $ to $[k^{2} \cdot c_i] \times [w_o \cdot h_o]$ (k is kernel window size) matrix for a more efficient matrix multiplication. \systemname{} divides the inputs across the channel dimension and processes the $im2col$ on maximum possible channels that can be processed at once.

For fully-connected layers, the main memory bottleneck is the parameters matrix $\mathcal{W}_{[O] \cdot [I]}$ that does not depend on the batch size. \systemname{} divides the matrix across the first dimension (rows), and processes the outputs on the maximum possible size of rows that fits inside the TEE for the corresponding layer.

%% file: tex_plots/mean_std_ref_acc.tex
\pgfplotsset{
    mean std attack figs style/.style={
        tick align=outside,
        tick pos=left,
        legend cell align={center},
        legend columns=2,
        legend style={%
        fill opacity=0.8, draw opacity=1, text opacity=1,at={(0.91,0.5)}, anchor=east,
        },
        scale only axis,
        width=.75\textwidth,
        height=0.3\textwidth,
        mark options={black},
        axis x line*=bottom,
        every non boxed x axis/.append style={x axis line style=-},
        every non boxed y axis/.append style={y axis line style=-},
        scaled y ticks = false,
        y tick label style={
        /pgf/number format/.cd,
            fixed,
            fixed zerofill,
            precision=2,
        /tikz/.cd
        },
    },
}
\begin{tikzpicture}
\begin{axis}[
mean std attack figs style,
xlabel={Epoch},
xmin=-9.95, xmax=208.95,
ylabel={Mean Accuracy},
ymin=0.238774999827147, ymax=0.958725011497736,
try min ticks=8,
axis y line*=left,
]

\addlegendimage{black,only marks,mark=o} %
\addlegendentry{lr=$10^{-1}$}

\addlegendimage{black,only marks,mark=square*} %
\addlegendentry{lr=$10^{-2}$}

\addlegendimage{black,line legend,solid}
\addlegendentry{mean}
\addlegendimage{black,line legend,dashed}
\addlegendentry{std}

\addplot [blue,mark=square*,mark repeat = 30,mark phase = 25]
table {%
0 0.476539969444275
1 0.548219919204712
2 0.611779928207397
3 0.649539947509766
4 0.664499998092651
5 0.700119972229004
6 0.725120067596436
7 0.733940005302429
8 0.726840019226074
9 0.751760005950928
10 0.759920001029968
11 0.774100065231323
12 0.764639973640442
13 0.786679983139038
14 0.771359920501709
15 0.798239946365356
16 0.787800073623657
17 0.79423999786377
18 0.808380007743835
19 0.805279970169067
20 0.800840020179749
21 0.808259963989258
22 0.807240009307861
23 0.824140071868896
24 0.821040034294128
25 0.819139957427979
26 0.802839994430542
27 0.821200013160706
28 0.826600074768066
29 0.816799998283386
30 0.839659929275513
31 0.831900000572205
32 0.832759976387024
33 0.828840017318726
34 0.845020055770874
35 0.832020044326782
36 0.832540035247803
37 0.822639942169189
38 0.842620015144348
39 0.839619994163513
40 0.84254002571106
41 0.834100008010864
42 0.846240043640137
43 0.835880041122437
44 0.844020009040833
45 0.840759992599487
46 0.8471999168396
47 0.854519963264465
48 0.841199994087219
49 0.848320007324219
50 0.891779899597168
51 0.893980026245117
52 0.893280029296875
53 0.89355993270874
55 0.89326000213623
56 0.894919991493225
57 0.894619941711426
58 0.895099997520447
60 0.894659996032715
61 0.895580053329468
62 0.893919944763184
63 0.894559979438782
64 0.895659923553467
65 0.894999980926514
66 0.893540024757385
67 0.896260023117065
68 0.895260095596313
69 0.894720077514648
71 0.895179986953735
72 0.894580006599426
73 0.895020008087158
74 0.894760012626648
75 0.895640015602112
76 0.894719958305359
77 0.894140005111694
79 0.894799947738647
81 0.895099997520447
82 0.893360018730164
83 0.894619941711426
85 0.895780086517334
86 0.894000053405762
87 0.894999980926514
88 0.894459962844849
89 0.8957200050354
91 0.896980047225952
95 0.89683997631073
96 0.897079944610596
97 0.897119998931885
98 0.897740006446838
100 0.896739959716797
101 0.897180080413818
102 0.896780014038086
103 0.898000001907349
104 0.897140026092529
105 0.89766001701355
106 0.896659970283508
107 0.897079944610596
109 0.897119998931885
110 0.897599935531616
111 0.897439956665039
112 0.897799968719482
113 0.896980047225952
114 0.897279977798462
115 0.897079944610596
116 0.897300004959106
117 0.898000001907349
118 0.897239923477173
119 0.89739990234375
120 0.897119998931885
121 0.898020029067993
122 0.897979974746704
123 0.898200035095215
124 0.897539973258972
126 0.897180080413818
127 0.898040056228638
129 0.897500038146973
130 0.89736008644104
131 0.89795994758606
132 0.897719979286194
133 0.89792001247406
134 0.897580027580261
135 0.897880077362061
136 0.897419929504395
138 0.898139953613281
139 0.898119926452637
140 0.897880077362061
141 0.898419976234436
143 0.897880077362061
144 0.898159980773926
145 0.897720098495483
146 0.898360013961792
147 0.898100018501282
148 0.898139953613281
149 0.897779941558838
150 0.897899985313416
151 0.89766001701355
153 0.897979974746704
154 0.89766001701355
156 0.898240089416504
157 0.898060083389282
158 0.898679971694946
159 0.89736008644104
160 0.898580074310303
161 0.898280024528503
163 0.89735996723175
164 0.897899985313416
165 0.897680044174194
166 0.898200035095215
167 0.897619962692261
168 0.898299932479858
169 0.897599935531616
170 0.898139953613281
171 0.898200035095215
172 0.897860050201416
173 0.898079991340637
174 0.897700071334839
175 0.898299932479858
176 0.898400068283081
177 0.897500038146973
179 0.898000001907349
180 0.897639989852905
181 0.898200035095215
182 0.898400068283081
183 0.897840023040771
184 0.898119926452637
185 0.89795994758606
186 0.898659944534302
187 0.897880077362061
188 0.898000001907349
189 0.897740006446838
191 0.897939920425415
194 0.89792001247406
195 0.898000001907349
196 0.898280024528503
198 0.89792001247406
199 0.898060083389282
};%

\addplot [red,mark=o,mark repeat = 30,mark phase = 10]
table {%
0 0.271499991416931
1 0.421640038490295
2 0.492980003356934
3 0.578019976615906
4 0.625339984893799
5 0.667420029640198
6 0.69894003868103
7 0.711220026016235
8 0.700940012931824
9 0.712219953536987
10 0.735599994659424
11 0.736000061035156
12 0.702859997749329
13 0.752740025520325
14 0.741440057754517
15 0.747499942779541
16 0.723479986190796
17 0.769160032272339
18 0.759999990463257
19 0.732959985733032
20 0.745800018310547
21 0.755100011825562
22 0.739279985427856
23 0.763400077819824
24 0.76338005065918
25 0.777220010757446
26 0.783240079879761
27 0.765779972076416
28 0.759799957275391
29 0.790620088577271
30 0.770900011062622
31 0.763419985771179
32 0.757459998130798
33 0.783779978752136
34 0.782959938049316
35 0.784620046615601
36 0.761520028114319
37 0.791939973831177
38 0.753620028495789
39 0.790740013122559
40 0.756259918212891
41 0.779160022735596
42 0.798020005226135
43 0.779860019683838
44 0.74370002746582
45 0.77347993850708
46 0.766579985618591
47 0.739580035209656
48 0.754179954528809
49 0.767660021781921
50 0.897939920425415
51 0.902080059051514
52 0.903959989547729
53 0.904299974441528
54 0.907520055770874
55 0.907079935073853
56 0.906960010528564
57 0.907340049743652
58 0.907279968261719
60 0.906219959259033
61 0.907060027122498
62 0.905699968338013
63 0.90637993812561
64 0.905079960823059
65 0.904439926147461
66 0.903059959411621
67 0.90034008026123
68 0.904039978981018
69 0.906019926071167
70 0.902920007705688
71 0.902159929275513
72 0.898560047149658
73 0.901419997215271
74 0.900840044021606
75 0.899459958076477
76 0.892879962921143
78 0.897799968719482
79 0.893980026245117
80 0.897000074386597
81 0.897000074386597
82 0.894079923629761
83 0.898659944534302
84 0.895599961280823
85 0.897239923477173
86 0.892699956893921
87 0.890820026397705
88 0.891680002212524
89 0.885319948196411
90 0.917979955673218
91 0.920720100402832
92 0.921259999275208
93 0.921479940414429
94 0.921200037002563
95 0.921819925308228
96 0.923239946365356
97 0.922039985656738
98 0.923179984092712
99 0.922659993171692
101 0.923179984092712
102 0.922640085220337
103 0.923500061035156
105 0.922879934310913
106 0.923140048980713
107 0.922559976577759
108 0.923619985580444
109 0.9235999584198
110 0.923160076141357
112 0.923500061035156
113 0.923280000686646
115 0.923939943313599
116 0.924080014228821
117 0.923259973526001
118 0.923480033874512
119 0.924360036849976
120 0.923680067062378
121 0.923720002174377
122 0.923460006713867
123 0.923759937286377
125 0.922999978065491
126 0.923520088195801
127 0.922240018844604
128 0.923259973526001
129 0.923160076141357
130 0.923579931259155
131 0.924220085144043
134 0.924560070037842
135 0.924040079116821
136 0.924700021743774
138 0.924360036849976
141 0.925019979476929
142 0.924579977989197
145 0.924819946289062
146 0.92467999458313
147 0.925559997558594
150 0.925119996070862
151 0.925459980964661
152 0.925999999046326
153 0.925679922103882
154 0.924799919128418
155 0.925300002098083
156 0.925580024719238
158 0.92493999004364
159 0.925519943237305
160 0.925260066986084
161 0.925260066986084
162 0.925699949264526
163 0.925279974937439
164 0.925140023231506
165 0.925539970397949
166 0.925459980964661
167 0.925140023231506
168 0.925440073013306
172 0.925480008125305
173 0.925740003585815
176 0.925300002098083
178 0.925100088119507
179 0.925639986991882
180 0.925019979476929
181 0.925420045852661
182 0.925459980964661
183 0.925260066986084
186 0.925420045852661
187 0.925240039825439
190 0.925620079040527
191 0.924839973449707
192 0.925400018692017
195 0.925179958343506
196 0.925800085067749
198 0.925279974937439
199 0.925519943237305
};%
\end{axis}
\begin{axis}[
mean std attack figs style,
axis y line*=right,
title={Mean and STD of reference accuracy across 5 repeats for each epoch},
xmin=-9.95, xmax=208.95,
ylabel={STD Accuracy},
ymin=-0.00315699560093745, ymax=0.0736067340798475,
ytick pos=right,
]

\addplot [dashed,red,mark=o,mark repeat = 30,mark phase = 20]
table {%
0 0.0266656875610352
1 0.018929123878479
2 0.021310567855835
3 0.0152480602264404
4 0.0137213468551636
5 0.0160665512084961
6 0.010479211807251
7 0.0163722038269043
8 0.0115429162979126
9 0.0267347097396851
10 0.0316838026046753
11 0.0115853548049927
12 0.0134123563766479
13 0.015558123588562
14 0.0366289615631104
15 0.0207773447036743
16 0.0224167108535767
17 0.0156800746917725
18 0.0137577056884766
19 0.0075995922088623
20 0.00864231586456299
21 0.0206289291381836
22 0.017494797706604
23 0.0146665573120117
24 0.0121806859970093
25 0.0171595811843872
26 0.0149894952774048
27 0.0214188098907471
28 0.0120837688446045
29 0.00788617134094238
30 0.0100885629653931
31 0.0150386095046997
32 0.00492525100708008
33 0.0149450302124023
34 0.00884139537811279
35 0.00978231430053711
36 0.0124229192733765
37 0.0249618291854858
38 0.00662517547607422
39 0.00910103321075439
40 0.0181947946548462
41 0.0176607370376587
42 0.00774943828582764
43 0.0209476947784424
44 0.0177942514419556
45 0.0090177059173584
46 0.00309574604034424
47 0.0112186670303345
48 0.0126376152038574
49 0.00392091274261475
50 0.00240278244018555
51 0.00213301181793213
52 0.00254273414611816
53 0.00249207019805908
54 0.00181949138641357
55 0.00105941295623779
56 0.00182473659515381
57 0.00128591060638428
58 0.00134909152984619
59 0.00222265720367432
60 0.00245320796966553
61 0.00196504592895508
62 0.00240945816040039
63 0.00179624557495117
64 0.00182271003723145
66 0.00225794315338135
67 0.00196635723114014
68 0.00113248825073242
69 0.00200533866882324
70 0.00195860862731934
71 0.00228512287139893
72 0.00247013568878174
73 0.00251185894012451
74 0.00175225734710693
75 0.0016942024230957
76 0.00132870674133301
77 0.00202834606170654
78 0.00223660469055176
79 0.00232040882110596
80 0.00205373764038086
81 0.00151658058166504
82 0.00172698497772217
83 0.00177919864654541
84 0.00198328495025635
85 0.00344753265380859
86 0.00196158885955811
87 0.00359773635864258
88 0.00401222705841064
89 0.00196611881256104
90 0.00250244140625
91 0.00228595733642578
92 0.0020449161529541
93 0.00229835510253906
95 0.00143885612487793
96 0.00207018852233887
97 0.00211882591247559
98 0.00157296657562256
99 0.00196325778961182
100 0.00215458869934082
101 0.00182366371154785
102 0.00225865840911865
103 0.00183188915252686
104 0.00166571140289307
105 0.00209248065948486
106 0.00221407413482666
107 0.001670241355896
108 0.00165700912475586
109 0.00224530696868896
110 0.00202882289886475
111 0.00210201740264893
112 0.00162351131439209
113 0.00227010250091553
114 0.00195491313934326
115 0.00204777717590332
116 0.00254011154174805
117 0.00168049335479736
119 0.00168168544769287
120 0.00138473510742188
121 0.00165581703186035
122 0.00209987163543701
124 0.00230014324188232
125 0.00214612483978271
126 0.00234472751617432
127 0.00245571136474609
128 0.00222837924957275
129 0.00253534317016602
130 0.00190949440002441
131 0.00213217735290527
132 0.00221478939056396
133 0.00180709362030029
134 0.00191247463226318
135 0.00207984447479248
136 0.00219404697418213
137 0.00215363502502441
139 0.00192296504974365
140 0.0018540620803833
141 0.00201332569122314
142 0.00200343132019043
143 0.00205850601196289
144 0.00194072723388672
145 0.0018925666809082
146 0.00188958644866943
147 0.00217533111572266
148 0.00198447704315186
149 0.00251829624176025
150 0.00205230712890625
151 0.0016094446182251
152 0.00201892852783203
153 0.00168454647064209
154 0.00252711772918701
155 0.00231051445007324
156 0.00191164016723633
157 0.00156664848327637
158 0.00171506404876709
159 0.00167524814605713
160 0.0015406608581543
161 0.00200939178466797
162 0.00223922729492188
163 0.00200259685516357
164 0.00245845317840576
165 0.00192701816558838
166 0.00206983089447021
167 0.00206923484802246
168 0.0016636848449707
169 0.00220632553100586
170 0.0016094446182251
171 0.00200700759887695
173 0.00190305709838867
174 0.00159871578216553
175 0.00208425521850586
176 0.00181436538696289
177 0.00197482109069824
178 0.00211000442504883
179 0.00221002101898193
180 0.00196433067321777
181 0.0018622875213623
182 0.00184500217437744
183 0.00209057331085205
185 0.0017012357711792
186 0.00182497501373291
187 0.00213861465454102
188 0.0019909143447876
189 0.00145959854125977
190 0.00190258026123047
191 0.00209724903106689
192 0.00166487693786621
193 0.00204980373382568
194 0.00240111351013184
195 0.00185143947601318
196 0.00183570384979248
197 0.0020519495010376
198 0.00235402584075928
199 0.00170838832855225
};%

\addplot [dashed,blue,mark=square*,mark repeat = 30,mark phase = 40]
table {%
0 0.0701174736022949
1 0.049144983291626
2 0.0485503673553467
3 0.0359764099121094
4 0.0340301990509033
5 0.0436633825302124
6 0.0222411155700684
7 0.0229049921035767
8 0.0250183343887329
9 0.0400348901748657
10 0.022229790687561
11 0.0195369720458984
12 0.0536820888519287
13 0.0124521255493164
14 0.029940128326416
15 0.0255625247955322
16 0.0275720357894897
17 0.014013409614563
18 0.0173821449279785
19 0.0294952392578125
20 0.0214520692825317
21 0.0250507593154907
22 0.038899302482605
23 0.0251545906066895
24 0.0280556678771973
25 0.0135695934295654
26 0.0340186357498169
27 0.015191912651062
28 0.0196863412857056
29 0.0180928707122803
30 0.0248122215270996
31 0.0181589126586914
32 0.0174640417098999
33 0.0327736139297485
34 0.0232404470443726
35 0.0231517553329468
36 0.0317033529281616
37 0.0202502012252808
38 0.0318804979324341
39 0.0275306701660156
40 0.0405147075653076
41 0.0278947353363037
42 0.0185952186584473
43 0.0132521390914917
44 0.0426353216171265
45 0.0215433835983276
46 0.0627932548522949
47 0.0279356241226196
48 0.0408192873001099
49 0.0216659307479858
50 0.00235509872436523
51 0.00267016887664795
52 0.00182163715362549
53 0.00155305862426758
54 0.00359416007995605
55 0.00236928462982178
56 0.00156402587890625
57 0.00331282615661621
58 0.00322639942169189
59 0.00213217735290527
60 0.00193953514099121
61 0.00303816795349121
62 0.00222980976104736
63 0.00204098224639893
64 0.00183022022247314
65 0.00111639499664307
66 0.00266575813293457
67 0.00549495220184326
68 0.0025554895401001
69 0.0017009973526001
70 0.00286388397216797
71 0.00440883636474609
72 0.00731158256530762
73 0.00234043598175049
74 0.00448262691497803
75 0.00353133678436279
76 0.00358355045318604
77 0.00462114810943604
78 0.00193703174591064
79 0.00508856773376465
80 0.00546479225158691
81 0.00391924381256104
82 0.00558257102966309
83 0.00597143173217773
84 0.00449573993682861
85 0.007973313331604
86 0.00485920906066895
87 0.00924217700958252
88 0.00563216209411621
89 0.00461888313293457
90 0.0019538402557373
91 0.00168097019195557
92 0.003104567527771
93 0.0023353099822998
94 0.00190365314483643
95 0.00235831737518311
96 0.0024254322052002
97 0.00206649303436279
98 0.00231635570526123
99 0.00159823894500732
100 0.0020594596862793
101 0.00197327136993408
102 0.00202047824859619
103 0.0026015043258667
104 0.00159001350402832
105 0.00226402282714844
106 0.00210297107696533
107 0.00169062614440918
108 0.00180160999298096
109 0.00160622596740723
110 0.00199759006500244
111 0.00119161605834961
112 0.00283336639404297
113 0.00196003913879395
114 0.00232160091400146
115 0.000725507736206055
116 0.00143027305603027
117 0.00233376026153564
118 0.00249588489532471
119 0.0020066499710083
120 0.00198733806610107
121 0.00142467021942139
122 0.00140082836151123
123 0.00275719165802002
124 0.00219869613647461
125 0.00183415412902832
126 0.00131511688232422
127 0.0016402006149292
128 0.000974893569946289
129 0.000332236289978027
130 0.000541806221008301
131 0.000820755958557129
132 0.00057065486907959
133 0.000691056251525879
134 0.000717282295227051
135 0.00057828426361084
136 0.000905513763427734
137 0.000786423683166504
138 0.000902414321899414
139 0.000939369201660156
140 0.00113773345947266
141 0.00098872184753418
142 0.00111961364746094
143 0.000765204429626465
144 0.000906825065612793
145 0.000667572021484375
146 0.000507593154907227
147 0.00113236904144287
148 0.000595331192016602
149 0.000985145568847656
150 0.000679373741149902
151 0.00105381011962891
152 0.000894427299499512
153 0.00075995922088623
154 0.00126969814300537
155 0.00123608112335205
156 0.000652432441711426
157 0.000847578048706055
158 0.00117909908294678
159 0.000851869583129883
160 0.00132906436920166
161 0.00104618072509766
162 0.00127279758453369
163 0.00109803676605225
164 0.00103843212127686
165 0.000801444053649902
166 0.00113773345947266
167 0.00138068199157715
168 0.00100922584533691
169 0.000997185707092285
170 0.000864863395690918
171 0.00077974796295166
172 0.00101470947265625
173 0.000886797904968262
174 0.00103461742401123
175 0.000386834144592285
176 0.00100994110107422
178 0.0010526180267334
179 0.000968694686889648
180 0.00135111808776855
181 0.000928163528442383
182 0.00112175941467285
183 0.000640630722045898
184 0.000733733177185059
185 0.00095832347869873
186 0.00121057033538818
187 0.000983119010925293
188 0.000899791717529297
189 0.000790953636169434
190 0.000998854637145996
191 0.00130021572113037
193 0.000749945640563965
194 0.00100910663604736
196 0.00117301940917969
197 0.0013573169708252
198 0.00123512744903564
199 0.00079345703125
};%

\end{axis}
\end{tikzpicture}%

%% file: tex_plots/best_gaps.tex
\pgfplotsset{
    best gap figs style/.style={
        tick align=outside,
        tick pos=left,
        legend cell align={center},
        legend columns=-1,
        legend style={
        fill opacity=0.8, draw opacity=1, text opacity=1,at={(.02,0.95)}, anchor=west,
        },
        scale only axis,
        width=.75\textwidth,
        height=0.3\textwidth,
        mark options={black},
        axis x line*=bottom,
        axis y line*=left,
        every non boxed x axis/.append style={x axis line style=-},
        every non boxed y axis/.append style={y axis line style=-},
        scaled y ticks = false,
        y tick label style={
        /pgf/number format/.cd,
            fixed,
            fixed zerofill,
            precision=2,
        /tikz/.cd
        },
        clip=false,
    },
}

\begin{tikzpicture}

\begin{axis}[
best gap figs style,
title={Highest test accuracy gap across 5 repeats},
xlabel={Epoch},
xmin=49, xmax=200,
ylabel={Best Difference},
ymin=-0.14,
ymax=0.05,
]

\addlegendimage{black,only marks,mark=o} %
\addlegendentry{lr=$10^{-1}$}

\addlegendimage{black,only marks,mark=square*} %
\addlegendentry{lr=$10^{-2}$}

\addlegendimage{black,line legend,solid}
\addlegendentry{clip=$5.10^{-4}$}
\addlegendimage{black,line legend,dashed}
\addlegendentry{clip=$10^{-3}$}

\addplot [red,mark=square*,mark repeat=30,mark phase=5,]
table {%
48 -0.0743999481201172
49 -0.0859999656677246
50 -0.110100030899048
51 -0.110399961471558
52 -0.110000014305115
53 -0.110199928283691
54 -0.109600067138672
56 -0.111700057983398
57 -0.111400008201599
58 -0.112499952316284
59 -0.107900023460388
60 -0.110599994659424
61 -0.110899925231934
62 -0.106299996376038
63 -0.108200073242188
64 -0.110399961471558
66 -0.105100035667419
67 -0.109300017356873
68 -0.108000040054321
69 -0.108399987220764
70 -0.108399987220764
71 -0.107500076293945
72 -0.109699964523315
73 -0.10699999332428
75 -0.107399940490723
76 -0.104099988937378
77 -0.105799913406372
78 -0.105299949645996
79 -0.104099988937378
80 -0.105999946594238
81 -0.106199979782104
82 -0.104300022125244
83 -0.104399919509888
84 -0.105499982833862
85 -0.105499982833862
86 -0.10319995880127
87 -0.104300022125244
88 -0.100399971008301
89 -0.104900002479553
90 -0.102999925613403
91 -0.10479998588562
93 -0.105499982833862
94 -0.105000019073486
95 -0.10479998588562
97 -0.10290002822876
98 -0.106299996376038
99 -0.107300043106079
100 -0.104900002479553
101 -0.106199979782104
102 -0.103800058364868
103 -0.106100082397461
104 -0.105799913406372
105 -0.103100061416626
106 -0.105400085449219
107 -0.104300022125244
108 -0.105299949645996
109 -0.104900002479553
110 -0.105999946594238
111 -0.103399991989136
112 -0.105000019073486
113 -0.10290002822876
114 -0.104599952697754
115 -0.103399991989136
116 -0.102499961853027
117 -0.104099988937378
118 -0.104900002479553
119 -0.105400085449219
120 -0.10669994354248
121 -0.106400012969971
122 -0.103100061416626
123 -0.104099988937378
124 -0.10450005531311
125 -0.10319995880127
126 -0.10509991645813
127 -0.105900049209595
128 -0.103600025177002
129 -0.105399966239929
130 -0.105299949645996
131 -0.105599999427795
132 -0.104699969291687
133 -0.104200005531311
134 -0.103299975395203
135 -0.104700088500977
136 -0.103100061416626
137 -0.104099988937378
138 -0.105499982833862
139 -0.104700088500977
140 -0.104099988937378
141 -0.105499982833862
142 -0.105599999427795
143 -0.103499889373779
144 -0.105000019073486
145 -0.104000091552734
146 -0.105900049209595
147 -0.104400038719177
148 -0.105599999427795
149 -0.106400012969971
150 -0.105900049209595
151 -0.104300022125244
152 -0.106100082397461
154 -0.103999972343445
156 -0.105700016021729
157 -0.107700109481812
158 -0.10509991645813
159 -0.105000019073486
161 -0.106500029563904
162 -0.105900049209595
163 -0.104900002479553
164 -0.104200005531311
165 -0.103300094604492
166 -0.106100082397461
167 -0.105700016021729
168 -0.105700016021729
169 -0.105900049209595
170 -0.104599952697754
171 -0.105200052261353
173 -0.104900002479553
174 -0.105200052261353
175 -0.107500076293945
177 -0.105000019073486
178 -0.105700016021729
180 -0.105399966239929
181 -0.105700016021729
182 -0.105299949645996
183 -0.105599999427795
184 -0.10450005531311
185 -0.103800058364868
186 -0.105599999427795
187 -0.105000019073486
188 -0.105299949645996
190 -0.10319995880127
191 -0.105599999427795
192 -0.105399966239929
193 -0.10669994354248
194 -0.105399966239929
195 -0.104900002479553
196 -0.10319995880127
197 -0.104200005531311
198 -0.104900002479553
199 -0.104099988937378
};
\addplot [red,dashed,mark=square*,mark repeat=30,mark phase=10,]
table {%
48 -0.0388998985290527
49 -0.0384999513626099
50 -0.0580999851226807
51 -0.0627000331878662
52 -0.0585000514984131
53 -0.0590999126434326
54 -0.0585000514984131
55 -0.0608999729156494
56 -0.0625
57 -0.0627000331878662
58 -0.0641000270843506
59 -0.0601999759674072
60 -0.0637999773025513
61 -0.0636999607086182
62 -0.0580999851226807
63 -0.0597000122070312
64 -0.0604000091552734
65 -0.0590999126434326
66 -0.0618000030517578
67 -0.065000057220459
68 -0.0612000226974487
69 -0.0595999956130981
71 -0.0619000196456909
72 -0.0604000091552734
73 -0.0601999759674072
74 -0.061500072479248
75 -0.0618999004364014
76 -0.0609999895095825
77 -0.0622999668121338
78 -0.0620999336242676
79 -0.0616999864578247
80 -0.0641999244689941
81 -0.0627000331878662
82 -0.0601000785827637
83 -0.0601999759674072
84 -0.062000036239624
85 -0.0599000453948975
86 -0.0608000755310059
87 -0.0592000484466553
88 -0.0562000274658203
89 -0.0640000104904175
90 -0.0618000030517578
91 -0.060499906539917
92 -0.0619000196456909
93 -0.0616999864578247
94 -0.0627000331878662
95 -0.063499927520752
96 -0.0627000331878662
97 -0.0627000331878662
98 -0.0631000995635986
99 -0.0623999834060669
100 -0.0609999895095825
101 -0.062999963760376
102 -0.0608999729156494
103 -0.0627000331878662
104 -0.063499927520752
105 -0.0634000301361084
106 -0.0611000061035156
107 -0.0629000663757324
108 -0.0611999034881592
109 -0.0618999004364014
110 -0.0631999969482422
111 -0.0634000301361084
113 -0.0629000663757324
114 -0.063499927520752
115 -0.0626000165939331
116 -0.0601999759674072
117 -0.0625
118 -0.0613000392913818
119 -0.0625
120 -0.0645999908447266
121 -0.0626000165939331
123 -0.0635000467300415
124 -0.0618000030517578
125 -0.063499927520752
126 -0.0627000331878662
127 -0.0623999834060669
128 -0.0622999668121338
129 -0.0613000392913818
130 -0.0626000165939331
132 -0.0626000165939331
133 -0.0632998943328857
134 -0.063499927520752
135 -0.0643000602722168
136 -0.0620999336242676
137 -0.0620999336242676
138 -0.063499927520752
139 -0.0633000135421753
140 -0.0629000663757324
141 -0.0629000663757324
142 -0.0640000104904175
143 -0.0634000301361084
144 -0.0631999969482422
145 -0.063499927520752
146 -0.0641000270843506
147 -0.0640000104904175
148 -0.0634000301361084
149 -0.0622999668121338
150 -0.062000036239624
151 -0.0638999938964844
152 -0.0632998943328857
153 -0.0634000301361084
154 -0.0615999698638916
155 -0.0627000331878662
157 -0.0633000135421753
158 -0.0643000602722168
159 -0.0629000663757324
160 -0.065000057220459
161 -0.0641999244689941
162 -0.0622999668121338
163 -0.0633000135421753
164 -0.0636999607086182
165 -0.0633000135421753
166 -0.0618999004364014
167 -0.0643999576568604
168 -0.0641000270843506
169 -0.0618000030517578
170 -0.0638000965118408
171 -0.0636999607086182
172 -0.0613999366760254
173 -0.0641999244689941
174 -0.0634000301361084
175 -0.0635000467300415
176 -0.0631999969482422
177 -0.0625998973846436
178 -0.0638999938964844
179 -0.0622999668121338
180 -0.062000036239624
181 -0.0631999969482422
182 -0.0634000301361084
183 -0.0637999773025513
184 -0.0636999607086182
185 -0.0631999969482422
187 -0.0638999938964844
188 -0.0629000663757324
189 -0.0638999938964844
190 -0.0641000270843506
191 -0.0627999305725098
192 -0.0628000497817993
193 -0.0638999938964844
194 -0.0633000135421753
195 -0.065000057220459
196 -0.0633000135421753
197 -0.0635000467300415
198 -0.0618999004364014
199 -0.0634000301361084
};
\addplot [blue,mark=o,mark repeat=30,mark phase=20,]
table {%
48 -0.0128999948501587
49 0.0379999876022339
50 -0.0149999856948853
52 -0.0093998908996582
53 -0.0136001110076904
54 -0.00479996204376221
55 -0.0125999450683594
56 -0.00839996337890625
57 -0.0115000009536743
58 -0.00629997253417969
59 -0.00629997253417969
60 -0.00480008125305176
61 -0.00960004329681396
62 -0.00550007820129395
63 -0.00440001487731934
64 -0.00639998912811279
65 -0.0055999755859375
66 -0.00240004062652588
67 0.00160002708435059
68 -0.00440001487731934
69 -0.00749993324279785
70 -0.00259995460510254
71 0
72 0.00769996643066406
73 -0.00290000438690186
74 0.00169992446899414
75 0.00119996070861816
76 0.00979995727539062
77 0.00309991836547852
78 0.00219988822937012
79 0.00429999828338623
80 0.00119996070861816
81 -0.000699996948242188
82 0.0046999454498291
83 -0.00150001049041748
84 0.00360000133514404
85 -0.00609993934631348
86 -0.00419998168945312
87 0.0130000114440918
88 0.0117000341415405
89 0.00940001010894775
90 -0.011699914932251
91 -0.0109000205993652
92 -0.00900006294250488
93 -0.010699987411499
94 -0.0111000537872314
95 -0.00999999046325684
96 -0.0130000114440918
97 -0.011199951171875
99 -0.0109000205993652
100 -0.012700080871582
101 -0.0131000280380249
102 -0.0130000114440918
103 -0.0112999677658081
104 -0.0104000568389893
105 -0.0121999979019165
106 -0.0109999179840088
107 -0.0108000040054321
108 -0.0123000144958496
109 -0.0112999677658081
110 -0.0125000476837158
111 -0.013200044631958
112 -0.00920009613037109
114 -0.0123000144958496
115 -0.013700008392334
116 -0.0136001110076904
117 -0.0104999542236328
119 -0.0125999450683594
120 -0.0118999481201172
121 -0.0115000009536743
122 -0.0144000053405762
123 -0.0117000341415405
124 -0.0112999677658081
125 -0.0127999782562256
126 -0.0148999691009521
127 -0.00919997692108154
128 -0.01419997215271
129 -0.0153000354766846
130 -0.0138000249862671
131 -0.0131000280380249
132 -0.0155000686645508
133 -0.0148000717163086
134 -0.0134000778198242
135 -0.0140999555587769
136 -0.0134000778198242
137 -0.0137999057769775
138 -0.0138000249862671
139 -0.0134000778198242
140 -0.0142999887466431
141 -0.0146000385284424
142 -0.0145000219345093
143 -0.0130000114440918
144 -0.0135999917984009
145 -0.0144000053405762
146 -0.0146999359130859
147 -0.0139999389648438
148 -0.0155000686645508
149 -0.0139000415802002
150 -0.0134000778198242
151 -0.0134000778198242
152 -0.016200065612793
153 -0.0151000022888184
154 -0.0131000280380249
155 -0.0124000310897827
156 -0.0162999629974365
158 -0.013200044631958
159 -0.0144999027252197
160 -0.0148999691009521
161 -0.0139999389648438
162 -0.0146999359130859
163 -0.0146999359130859
164 -0.0148999691009521
165 -0.0146000385284424
166 -0.0135999917984009
167 -0.013200044631958
169 -0.0149999856948853
170 -0.0153000354766846
171 -0.0139999389648438
172 -0.0135999917984009
173 -0.0155999660491943
174 -0.0146999359130859
175 -0.0156999826431274
176 -0.0144000053405762
177 -0.0154000520706177
178 -0.0134000778198242
179 -0.0151000022888184
180 -0.013700008392334
181 -0.0142999887466431
182 -0.012700080871582
183 -0.0148999691009521
184 -0.0151998996734619
185 -0.0132999420166016
186 -0.0137999057769775
187 -0.0148999691009521
188 -0.0138000249862671
190 -0.0148999691009521
191 -0.0146999359130859
192 -0.013200044631958
193 -0.0144000053405762
194 -0.0148999691009521
195 -0.0131000280380249
196 -0.0147000551223755
198 -0.0135999917984009
199 -0.0146000385284424
};
\addplot [blue,dashed,mark=o,mark repeat=30,mark phase=5,]
table {%
48 0.0638999938964844
49 -0.000400066375732422
50 -0.00600004196166992
51 -0.00920009613037109
52 -0.0055999755859375
53 -0.00360000133514404
54 -0.00789999961853027
55 -0.00419998168945312
56 -0.00320005416870117
57 -0.0037999153137207
58 -0.00219988822937012
59 0.000599980354309082
60 -0.00199997425079346
61 0.000699996948242188
62 9.98973846435547e-05
63 -0.00329995155334473
64 -0.00250005722045898
65 -0.00279998779296875
66 -0.000999927520751953
67 0.00510001182556152
68 -0.00259995460510254
69 -0.00530004501342773
70 0.000800013542175293
72 0.00959992408752441
73 0.00169992446899414
74 0.00530004501342773
75 0.00730001926422119
76 0.0101000070571899
77 0.0104000568389893
78 0.00320005416870117
79 0.0149999856948853
80 0.00160002708435059
81 0.00720000267028809
82 0.00859999656677246
83 0.00690007209777832
84 0.00349998474121094
85 0.000800013542175293
86 0.00489997863769531
87 0.0181999206542969
88 0.0103000402450562
89 0.0120000839233398
90 -0.000699996948242188
91 -0.00209999084472656
92 -0.00130009651184082
93 -0.00239992141723633
94 -0.00220000743865967
95 -0.00250005722045898
96 -0.00370001792907715
97 -0.00519990921020508
98 -0.00609993934631348
99 -0.006600022315979
100 -0.00499999523162842
101 -0.00569999217987061
102 -0.00479996204376221
103 -0.00339996814727783
104 -0.00390005111694336
105 -0.0037999153137207
106 -0.00639998912811279
107 -0.00600004196166992
108 -0.00499999523162842
109 -0.00670003890991211
110 -0.006600022315979
111 -0.00609993934631348
112 -0.0074000358581543
113 -0.00449991226196289
114 -0.00230002403259277
115 -0.0055999755859375
116 -0.00429999828338623
117 -0.00410008430480957
118 -0.00340008735656738
119 -0.00410008430480957
120 -0.00699996948242188
121 -0.00349998474121094
122 -0.00730001926422119
123 -0.00460004806518555
124 -0.00239992141723633
125 -0.00329995155334473
126 -0.00440001487731934
127 -0.00239992141723633
128 -0.00489997863769531
129 -0.00620007514953613
130 -0.00410008430480957
131 -0.0046999454498291
132 -0.00580000877380371
133 -0.00600004196166992
134 -0.00590002536773682
135 -0.00629997253417969
136 -0.00629997253417969
137 -0.00589990615844727
138 -0.00460004806518555
139 -0.00600004196166992
140 -0.00580000877380371
141 -0.00489997863769531
142 -0.00510001182556152
143 -0.00550007820129395
144 -0.00479996204376221
146 -0.0046999454498291
147 -0.00499999523162842
148 -0.00440001487731934
149 -0.00480008125305176
150 -0.00460004806518555
151 -0.00609993934631348
152 -0.00539994239807129
154 -0.00730001926422119
155 -0.00580000877380371
156 -0.00729990005493164
157 -0.00460004806518555
158 -0.00609993934631348
159 -0.00659990310668945
160 -0.00600004196166992
161 -0.0055999755859375
162 -0.00590002536773682
163 -0.00479996204376221
164 -0.00679993629455566
165 -0.00670003890991211
166 -0.00590002536773682
167 -0.00629997253417969
168 -0.00629997253417969
169 -0.00609993934631348
170 -0.00550007820129395
171 -0.00620007514953613
172 -0.00539994239807129
173 -0.0055999755859375
174 -0.00480008125305176
175 -0.00580000877380371
176 -0.00419998168945312
177 -0.00619995594024658
178 -0.0055999755859375
180 -0.00670003890991211
181 -0.00619995594024658
182 -0.00629997253417969
183 -0.00600004196166992
184 -0.00619995594024658
185 -0.00499999523162842
186 -0.0046999454498291
187 -0.00619995594024658
188 -0.00589990615844727
189 -0.00580000877380371
190 -0.00679993629455566
191 -0.00489997863769531
192 -0.00569999217987061
193 -0.00709998607635498
194 -0.00659990310668945
195 -0.00550007820129395
196 -0.00580000877380371
197 -0.006600022315979
198 -0.00549995899200439
199 -0.00620007514953613
};
\end{axis}
\end{tikzpicture}%

%% file: tex_plots/worst_gaps.tex
\pgfplotsset{
    worst gap figs style/.style={
        tick align=outside,
        tick pos=left,
        legend cell align={center},
        legend columns=-1,
        legend style={
        fill opacity=0.8, draw opacity=1, text opacity=1,at={(.05,0.95)}, anchor=west,
        },
        scale only axis,
        width=.75\textwidth,
        height=0.3\textwidth,
        mark options={black},
        axis x line*=bottom,
        axis y line*=left,
        every non boxed x axis/.append style={x axis line style=-},
        every non boxed y axis/.append style={y axis line style=-},
        scaled y ticks = false,
        y tick label style={
        /pgf/number format/.cd,
            fixed,
            fixed zerofill,
            precision=2,
        /tikz/.cd
        },
        clip=false,
    },
}
\begin{tikzpicture}
\begin{axis}[
worst gap figs style,
title={Lowest test accuracy gap across 5 repeats},
xlabel={Epoch},
xmin=49, xmax=200,
ylabel={Worst Difference},
ymin=-0.15, ymax=0.051,
]
\addplot [red,mark=square*,mark repeat=30,mark phase=5,]
table {%
49 -0.109200000762939
50 -0.118399977684021
51 -0.120100021362305
52 -0.117599964141846
53 -0.120300054550171
54 -0.117799997329712
55 -0.119499921798706
56 -0.119300007820129
57 -0.118299961090088
58 -0.120499968528748
59 -0.11899995803833
60 -0.118299961090088
61 -0.120199918746948
62 -0.115499973297119
63 -0.120200037956238
64 -0.121600031852722
65 -0.115100026130676
66 -0.12090003490448
67 -0.118399977684021
68 -0.116100072860718
69 -0.117199897766113
70 -0.11710000038147
71 -0.117799997329712
72 -0.115999937057495
73 -0.120100021362305
74 -0.115800023078918
75 -0.117799997329712
76 -0.119099974632263
78 -0.113899946212769
79 -0.116899967193604
80 -0.117900013923645
81 -0.115999937057495
82 -0.114400029182434
83 -0.117399930953979
84 -0.115700006484985
85 -0.118000030517578
86 -0.115200042724609
87 -0.114099979400635
88 -0.115399956703186
89 -0.114099979400635
90 -0.11489999294281
91 -0.116699934005737
92 -0.114700078964233
93 -0.115700006484985
94 -0.11489999294281
95 -0.113899946212769
97 -0.116199970245361
98 -0.115900039672852
99 -0.116999983787537
100 -0.113399982452393
101 -0.116099953651428
102 -0.113300085067749
103 -0.115800023078918
104 -0.114400029182434
105 -0.116899967193604
106 -0.115599989891052
107 -0.114000082015991
108 -0.115100026130676
109 -0.115200042724609
110 -0.116999983787537
111 -0.116899967193604
112 -0.115399956703186
113 -0.116199970245361
114 -0.114399909973145
115 -0.115200042724609
116 -0.117300033569336
117 -0.115000009536743
118 -0.113700032234192
119 -0.115499973297119
120 -0.114199995994568
121 -0.115400075912476
122 -0.117799997329712
123 -0.117500066757202
124 -0.115499973297119
125 -0.115900039672852
126 -0.115700006484985
127 -0.116600036621094
128 -0.115400075912476
129 -0.115200042724609
130 -0.115400075912476
131 -0.115200042724609
132 -0.116099953651428
133 -0.116499900817871
134 -0.115499973297119
135 -0.116300106048584
136 -0.115599989891052
137 -0.116400003433228
138 -0.115400075912476
139 -0.116300106048584
140 -0.116500020027161
141 -0.11710000038147
142 -0.115900039672852
143 -0.115700006484985
144 -0.116500020027161
145 -0.115100026130676
146 -0.115200042724609
147 -0.117500066757202
148 -0.115099906921387
149 -0.115599989891052
150 -0.11710000038147
151 -0.115299940109253
152 -0.114000082015991
153 -0.114399909973145
154 -0.11680006980896
155 -0.11680006980896
156 -0.114300012588501
157 -0.115700006484985
158 -0.116299986839294
160 -0.115799903869629
161 -0.116600036621094
162 -0.117799997329712
163 -0.113399982452393
164 -0.118799924850464
165 -0.116199970245361
166 -0.114700078964233
167 -0.116100072860718
168 -0.115800023078918
169 -0.115000009536743
170 -0.115999937057495
171 -0.116300106048584
172 -0.114799976348877
173 -0.116500020027161
174 -0.115400075912476
176 -0.115999937057495
177 -0.11680006980896
180 -0.114700078964233
181 -0.116400003433228
182 -0.117900013923645
183 -0.115400075912476
184 -0.116199970245361
185 -0.117199897766113
186 -0.117599964141846
187 -0.116400003433228
188 -0.115800023078918
189 -0.114500045776367
190 -0.116600036621094
191 -0.117899894714355
192 -0.114199995994568
193 -0.118000030517578
194 -0.117899894714355
195 -0.114799976348877
196 -0.115999937057495
197 -0.115299940109253
198 -0.115000009536743
199 -0.114500045776367
};
\addplot [red,dashed,mark=square*,mark repeat=30,mark phase=5,]
table {%
49 -0.0580000877380371
50 -0.0716999769210815
51 -0.0694999694824219
52 -0.0706000328063965
53 -0.0710999965667725
54 -0.069100022315979
55 -0.0697000026702881
56 -0.07069993019104
57 -0.0706000328063965
58 -0.0709999799728394
59 -0.0708999633789062
60 -0.0713000297546387
61 -0.076200008392334
62 -0.072700023651123
63 -0.0715999603271484
64 -0.0713000297546387
65 -0.0699999332427979
66 -0.073699951171875
67 -0.0712000131607056
68 -0.072700023651123
69 -0.0713000297546387
70 -0.0755999088287354
71 -0.0722000598907471
72 -0.0720999240875244
73 -0.073699951171875
74 -0.0716999769210815
75 -0.073199987411499
76 -0.0729999542236328
77 -0.0724999904632568
78 -0.0713999271392822
79 -0.0752999782562256
80 -0.0729000568389893
81 -0.073699951171875
82 -0.0710000991821289
83 -0.0706000328063965
84 -0.0724999904632568
85 -0.0727999210357666
86 -0.0715000629425049
87 -0.0730999708175659
88 -0.0724999904632568
89 -0.0709999799728394
90 -0.0746999979019165
91 -0.0729000568389893
92 -0.0715999603271484
93 -0.0731000900268555
94 -0.0717999935150146
95 -0.07069993019104
96 -0.0728000402450562
99 -0.0734000205993652
100 -0.0724999904632568
101 -0.0709999799728394
102 -0.0738000869750977
103 -0.0726000070571899
105 -0.0731000900268555
106 -0.0727999210357666
107 -0.0715999603271484
108 -0.072700023651123
109 -0.0723999738693237
110 -0.0723999738693237
111 -0.0717999935150146
112 -0.0728000402450562
113 -0.0729999542236328
114 -0.0729999542236328
115 -0.0719000101089478
116 -0.0736000537872314
117 -0.0731000900268555
118 -0.0724000930786133
119 -0.0722999572753906
120 -0.0715000629425049
121 -0.0720000267028809
123 -0.0750999450683594
124 -0.0734000205993652
125 -0.0738000869750977
126 -0.0727999210357666
127 -0.0746999979019165
128 -0.0743000507354736
129 -0.0713000297546387
130 -0.0719000101089478
131 -0.0736000537872314
132 -0.0736000537872314
133 -0.0724999904632568
134 -0.0734000205993652
135 -0.0728000402450562
136 -0.0743000507354736
137 -0.0737999677658081
138 -0.0735000371932983
139 -0.0722999572753906
140 -0.0736000537872314
141 -0.0740000009536743
142 -0.0726000070571899
143 -0.0730999708175659
144 -0.0720000267028809
145 -0.0738999843597412
146 -0.073199987411499
147 -0.0738000869750977
148 -0.073699951171875
149 -0.0738999843597412
150 -0.0738999843597412
151 -0.0731000900268555
152 -0.0734000205993652
153 -0.0727999210357666
154 -0.0733000040054321
155 -0.0744999647140503
156 -0.0738999843597412
157 -0.0723999738693237
158 -0.0734000205993652
159 -0.0715000629425049
161 -0.0734000205993652
162 -0.0734000205993652
163 -0.0715000629425049
164 -0.0724999904632568
165 -0.0720000267028809
166 -0.0738000869750977
167 -0.0715999603271484
168 -0.0736000537872314
169 -0.0738000869750977
170 -0.0726000070571899
171 -0.0729999542236328
172 -0.0736000537872314
173 -0.072700023651123
174 -0.0726000070571899
175 -0.0743999481201172
176 -0.0729999542236328
177 -0.0736000537872314
178 -0.0724999904632568
179 -0.0733000040054321
180 -0.0720000267028809
181 -0.0742000341415405
182 -0.073199987411499
183 -0.0729000568389893
184 -0.0719000101089478
185 -0.072700023651123
186 -0.0744999647140503
187 -0.0722999572753906
188 -0.072700023651123
189 -0.0740000009536743
190 -0.073699951171875
191 -0.0724999904632568
192 -0.0729000568389893
193 -0.0712000131607056
194 -0.0733000040054321
195 -0.0724000930786133
196 -0.0734000205993652
197 -0.0727999210357666
198 -0.0738999843597412
199 -0.0736000537872314
};
\addplot [blue,mark=o,mark repeat=30,mark phase=15,]
table {%
49 -0.0973000526428223
50 -0.0202999114990234
51 -0.0204000473022461
52 -0.0214999914169312
53 -0.0152000188827515
54 -0.021399974822998
55 -0.0147000551223755
56 -0.0178999900817871
57 -0.0168000459671021
58 -0.0209000110626221
59 -0.0141000747680664
60 -0.0166000127792358
61 -0.0199999809265137
62 -0.01419997215271
63 -0.0124000310897827
64 -0.0200999975204468
65 -0.00909996032714844
66 -0.0116000175476074
67 -0.0142999887466431
68 -0.0148000717163086
69 -0.0211000442504883
70 -0.0141000747680664
71 -0.0188999176025391
72 -0.0153999328613281
73 -0.00999999046325684
74 -0.0151000022888184
75 -0.00919997692108154
76 -0.0148000717163086
77 -0.0139000415802002
78 -0.00919997692108154
79 -0.00970005989074707
80 -0.0490000247955322
81 -0.0288000106811523
82 -0.0174000263214111
83 -0.0169000625610352
84 -0.0123000144958496
85 -0.0199000835418701
86 -0.0153999328613281
87 -0.00839996337890625
88 -0.0197999477386475
89 -0.00890004634857178
90 -0.017799973487854
91 -0.0178999900817871
92 -0.0194000005722046
93 -0.0167000293731689
94 -0.0169000625610352
95 -0.0178999900817871
96 -0.0185999870300293
97 -0.0167000293731689
98 -0.0211999416351318
99 -0.0187000036239624
100 -0.0180000066757202
101 -0.0183000564575195
102 -0.0174999237060547
103 -0.0204000473022461
104 -0.0180000066757202
105 -0.0192999839782715
106 -0.0192999839782715
107 -0.0190999507904053
108 -0.0176999568939209
110 -0.0194000005722046
111 -0.0187000036239624
112 -0.0195999145507812
113 -0.0185000896453857
114 -0.0192000865936279
115 -0.0160999298095703
116 -0.0171999931335449
117 -0.0188000202178955
118 -0.0194000005722046
119 -0.0216000080108643
120 -0.0210000276565552
121 -0.0187000036239624
122 -0.0181999206542969
123 -0.0188999176025391
124 -0.0187000036239624
125 -0.0178000926971436
126 -0.0188999176025391
127 -0.0153000354766846
128 -0.0162999629974365
129 -0.0170999765396118
130 -0.016200065612793
131 -0.0169999599456787
132 -0.0166000127792358
133 -0.0176000595092773
134 -0.017300009727478
135 -0.0167000293731689
136 -0.0180000066757202
137 -0.0176999568939209
138 -0.0169000625610352
139 -0.0188000202178955
140 -0.0170999765396118
141 -0.017799973487854
142 -0.0169999599456787
143 -0.0178999900817871
144 -0.0181999206542969
145 -0.0169999599456787
146 -0.0174000263214111
147 -0.0188999176025391
148 -0.0178999900817871
149 -0.0176999568939209
150 -0.0176999568939209
151 -0.0188000202178955
152 -0.0190000534057617
153 -0.0180000066757202
154 -0.0185000896453857
155 -0.0180000066757202
156 -0.0182000398635864
157 -0.0181000232696533
158 -0.0174999237060547
159 -0.0194000005722046
161 -0.0183000564575195
162 -0.0190000534057617
163 -0.0181000232696533
164 -0.0183000564575195
165 -0.0188000202178955
166 -0.0195000171661377
167 -0.0185999870300293
168 -0.0185999870300293
169 -0.0181999206542969
170 -0.0189000368118286
171 -0.0181000232696533
173 -0.0183999538421631
174 -0.0190999507904053
175 -0.0169999599456787
178 -0.0175000429153442
179 -0.0174000263214111
180 -0.0187000036239624
181 -0.0176999568939209
182 -0.0185999870300293
184 -0.0176000595092773
185 -0.0187000036239624
186 -0.0190000534057617
187 -0.017799973487854
188 -0.0192000865936279
189 -0.0187000036239624
191 -0.0182000398635864
193 -0.0190999507904053
194 -0.0192000865936279
195 -0.0181000232696533
196 -0.0185999870300293
197 -0.017799973487854
198 -0.0180000066757202
199 -0.0195000171661377
};
\addplot [blue,dashed,mark=o,mark repeat=30,mark phase=20,]
table {%
49 -0.108599901199341
50 -0.0166000127792358
51 -0.0157999992370605
52 -0.0130999088287354
53 -0.0131000280380249
54 -0.0155999660491943
55 -0.012700080871582
56 -0.0146000385284424
58 -0.0128999948501587
59 -0.00900006294250488
61 -0.0139999389648438
62 -0.00960004329681396
63 -0.016200065612793
64 -0.00789999961853027
65 -0.00999999046325684
66 -0.00720000267028809
67 -0.00850009918212891
68 -0.013700008392334
69 -0.00940001010894775
70 -0.00670003890991211
71 -0.00729990005493164
72 -0.0074000358581543
73 -0.00839996337890625
74 -0.00639998912811279
75 -0.00399994850158691
76 -0.00519990921020508
77 -0.00390005111694336
78 -0.00300002098083496
79 -0.00279998779296875
80 -0.0185999870300293
81 -0.00720000267028809
82 -0.0135999917984009
83 -0.0151000022888184
84 -0.00129997730255127
85 -0.0174999237060547
86 -0.00620007514953613
87 -0.0131000280380249
88 -0.00160002708435059
89 -0.00520002841949463
90 -0.0128999948501587
91 -0.0145000219345093
92 -0.0135999917984009
93 -0.0134000778198242
94 -0.0121999979019165
95 -0.0120999813079834
96 -0.0128999948501587
97 -0.011199951171875
98 -0.0160999298095703
99 -0.0121999979019165
100 -0.0128999948501587
101 -0.013200044631958
102 -0.0123000144958496
103 -0.0144000053405762
104 -0.0115000009536743
105 -0.0130000114440918
106 -0.0115000009536743
107 -0.010200023651123
108 -0.0127999782562256
109 -0.0134999752044678
111 -0.013200044631958
112 -0.0169000625610352
113 -0.01419997215271
114 -0.0123000144958496
115 -0.0134000778198242
116 -0.0134999752044678
117 -0.0139000415802002
118 -0.0148999691009521
119 -0.013700008392334
121 -0.013200044631958
123 -0.0134999752044678
124 -0.01419997215271
125 -0.015700101852417
126 -0.013200044631958
127 -0.0111000537872314
128 -0.0118000507354736
129 -0.0113999843597412
130 -0.0115000009536743
131 -0.0124000310897827
132 -0.0116000175476074
133 -0.0115000009536743
134 -0.0104000568389893
135 -0.0109999179840088
136 -0.011199951171875
137 -0.0104999542236328
138 -0.0128999948501587
139 -0.0108000040054321
140 -0.0131000280380249
141 -0.0121999979019165
142 -0.0132999420166016
143 -0.0118999481201172
144 -0.0118000507354736
145 -0.011199951171875
146 -0.0121999979019165
147 -0.0125999450683594
148 -0.0126999616622925
149 -0.0132999420166016
150 -0.0116000175476074
152 -0.0132999420166016
153 -0.0121999979019165
154 -0.0127999782562256
155 -0.0128999948501587
157 -0.0110000371932983
158 -0.0118999481201172
159 -0.0125000476837158
160 -0.0120999813079834
161 -0.0130000114440918
162 -0.012700080871582
163 -0.0105999708175659
164 -0.011699914932251
165 -0.011199951171875
166 -0.0120000839233398
167 -0.0118999481201172
168 -0.0116000175476074
169 -0.0126999616622925
170 -0.0112999677658081
172 -0.0121999979019165
173 -0.0116000175476074
174 -0.0120999813079834
175 -0.0115000009536743
176 -0.0120000839233398
177 -0.0109999179840088
178 -0.0115000009536743
179 -0.0109999179840088
180 -0.0123000144958496
181 -0.0118000507354736
182 -0.0120999813079834
183 -0.0131000280380249
184 -0.0109000205993652
185 -0.0125999450683594
186 -0.0123999118804932
187 -0.0115000009536743
188 -0.0121999979019165
189 -0.0125999450683594
190 -0.0116000175476074
191 -0.0120999813079834
192 -0.0123000144958496
193 -0.0118999481201172
194 -0.011199951171875
195 -0.0109999179840088
196 -0.0117000341415405
197 -0.011699914932251
198 -0.0111000537872314
199 -0.0120000839233398
};
\end{axis}
\end{tikzpicture}%